\documentclass[ijoc, nonblindrev]{informs3}

\OneAndAHalfSpacedXI

\usepackage{graphicx}
\usepackage{flafter,placeins,amsmath, amsfonts, amssymb, hyperref, epstopdf, stmaryrd}
\usepackage[normalem]{ulem}
\usepackage{booktabs}
\usepackage[dvipsnames]{xcolor}
\usepackage{tikz}
\usetikzlibrary{positioning}
\usepackage{todonotes}
\usepackage{bbm}
\usetikzlibrary{shapes,snakes}
\usepackage{algorithmic}

\usepackage{makecell}
\usepackage{multirow}

\usepackage{caption}
              
\usepackage{boxhandler}

\usepackage{natbib}
 \bibpunct[, ]{(}{)}{,}{a}{}{,}%

\usepackage[vlined, ruled]{algorithm2e}
\usepackage{subcaption}
\usepackage{adjustbox}

\usepackage{makecell} 

\usepackage{color}
\usepackage{float}
\newcommand{\rev}[1]{{\color{black} #1}}
\newcommand{\revcs}[1]{{\color{black} #1}}

\newcommand\T{\rule{0pt}{2.6ex}}       

\newtheorem{observation}{Observation}
\theoremstyle{definition}
\newtheorem{exmp}{Example}[section]
\TheoremsNumberedThrough     
\ECRepeatTheorems

\EquationsNumberedThrough    

\MANUSCRIPTNO{JOC-2020-09-OA-218.R1}

\begin{document}


\RUNAUTHOR{Jena et al.}

\RUNTITLE{ On the estimation of discrete choice models to capture irrational customer behaviors}

\TITLE{ On the estimation of discrete choice models to capture irrational customer behaviors
}

\ARTICLEAUTHORS{%

\AUTHOR{Sanjay Dominik Jena}
\AFF{\'Ecole des Sciences de la Gestion, Universit\'e du Qu\'ebec \`a Montr\'eal,\\ Centre interuniversitaire de recherche sur les r\'eseaux d'entreprise, la logistique et le transport (CIRRELT)
\EMAIL{jena.sanjay-dominik@uqam.ca}} 

\AUTHOR{Andrea Lodi}
\AFF{Canada Excellence Research Chair in Data-Science for Real-time Decision-Making,\\Polytechnique Montr\'eal,\\
\EMAIL{andrea.lodi@polymtl.ca}}

\AUTHOR{Claudio Sole}
\AFF{Canada Excellence Research Chair in Data-Science for Real-time Decision-Making, \\Polytechnique Montr\'eal,\\
\EMAIL{claudio.sole@polymtl.ca}}

} 

\ABSTRACT{%
\revcs{The\textit{ Random Utility Maximization} model is by far the most adopted framework to estimate consumer choice behavior. However, behavioral economics has provided strong empirical evidence of \textit{irrational} choice behavior, such as \textit{halo} effects, that are incompatible with this framework. Models belonging to the Random Utility Maximization family may therefore not accurately capture such irrational behavior.} 
Hence, more general choice models, overcoming such limitations, have been proposed. \textcolor{black}{However, the flexibility of such models comes at the price of increased risk of overfitting. As such, estimating such models remains a challenge.}
In this work, we propose an estimation method for the recently proposed Generalized Stochastic Preference choice model, which subsumes the family of Random Utility Maximization models and is capable of capturing halo effects. Specifically, we show how to use partially-ranked preferences to efficiently model rational and irrational customer types from transaction data. Our estimation procedure is based on column generation, where relevant customer types are efficiently extracted by expanding a tree-like data structure containing the customer behaviors. \revcs{Further, we propose a new dominance rule among customer types whose effect is to prioritize low orders of interactions among products.} An extensive set of experiments assesses the predictive accuracy of the proposed approach. 
\rev{Our results show that accounting for irrational preferences can boost predictive accuracy by 12.5\% on average, when tested on a real-world dataset from a large chain of grocery and drug stores.}
 
}%


\maketitle


\section{Introduction}
\label{sec:Intro}
\revcs{Accurately forecasting the demand of certain products or services is of crucial importance in the context of supply-chain optimization and retail operations.} Most often, a predictive model must be learned from historical data representing the choice behavior of an agent faced with a discrete set of alternatives, called the \textit{offer set}. 
A common assumption when dealing with demand estimation is to consider product demands as independent from each other, resulting in the independent demand model \citep[see, e.g.,][]{strauss2018review, Talluri2004a}. However, it is well known that this assumption does not hold in many real-life scenarios and that product demands interact through substitution and halo effects. In general, we consider alternative $x$ a substitute of alternative $y$ if the presence of $x$ in the offer set decreases the probability of $y$ being chosen. On the contrary, we refer to an halo effect if the presence of $x$ in the offer set increases the attractiveness of $y$, and thus its likelihood of being chosen.

Discrete choice models have been widely adopted to model substitution. Among them, the family of choice models that received the most attention in the literature is undoubtedly the one of \textit{Random Utility Maximization} (RUM) models \citep{thurstone1927RUM, block1959random, Luce1959}. Choice models belonging to the RUM  family assume that a random utility is assigned to every alternative. Utilities are modeled as random variables, and different choices about their distribution lead to different choice models. When faced with an offer set, the decision maker samples a vector of utilities and picks the option with the highest one, so as to maximize her expected payoff.  

 \revcs{The standard theory of rational choice assumes the relative preference between two alternatives does not depend on the other products in the offer set. Hence, if alternative $x$ is preferred to alternative $y$ in a given offer set $S$, the same should hold for any other offer set $S' \neq S$. Starting from this assumption, known as \textit{Independence of Irrelevant Alternatives} (IIA), \cite{Luce1959} derived the Multinomial Logit (MNL) model, which is arguably the most famous RUM choice model.} Its popularity stems from the facts that it can be efficiently estimated, it is interpretable and, when used for decision making, it allows to benefit from appealing theoretical and computational properties. The Multinomial Logit model lacks, however, flexibility\revcs{, imposing specific patterns of substitutions among alternatives. In particular, the logit formula implies that the ratio between the choice probabilities of two alternatives does not depend on the other (irrelevant) alternatives in the offer set. In response, many models have been proposed to overcome such limitations, so as to capture more complex patterns of substitutions. In the Nested Logit model, for example, this is achieved by assuming IIA holds only among groups of similar alternatives. In the Mixed Multinomial logit (MMNL) and Rank-Based \citep{farias13} choice models, instead, violations of the IIA assumption are captured by aggregating several IIA-consistent classes of customers. Notably, some of these models, such as the MMNL, the Rank-Based and the Markov Chain \citep{Blanchet2016} choice models, can theoretically approximate any RUM choice model and therefore capture arbitrarily complex substitution effects. We refer the interested reader to the computational study of \cite{comparativeBerbeglia} for a deeper overview on RUM choice models and their generalization performances.}




All models that belong to the RUM family obey the so-called \textit{Regularity} assumption, which states that the introduction of an option in the offer set cannot increase the probability of another alternative being chosen. Hence, they cannot capture halo effects. Nevertheless, many studies in the literature of behavioral economics corroborated the reproducibility and robustness of this type of choice behaviors \citep[see, e.g.,][]{attreffect1_simonson1989choice, attreffect2_huber1982adding}, incompatible with the theory of utility maximization and therefore referred to as \textit{irrational}. 
\revcs{
In the remainder of this paper, we therefore refer to rational behavior as one that can be captured by RUM models, and to irrational behavior as one that cannot.
In order to better illustrate  the kind of choice scenario in which violations of the Regularity assumption may arise, we report below  the results of a choice experiment from the seminal work of \cite{simonson1992choice_exp}.}

\vspace{0.5cm}

\begin{exmp}[Choice Experiment from \cite{simonson1992choice_exp}]
\label{example}
\revcs{
In this experiment, respondents were asked to choose among three camera models, which differ in terms of price and quality. Specifically, the three models were (1) a Minolta X-370 camera, priced at \$170, (2) a Minolta 3000i, priced at \$240 and (3) a Minolta 7000i, with a price of \$470. Table \ref{tab:MarketShare} reports the market shares of the alternatives in the choice scenarios $S_1$, where only options \{1,2\} are offered, and $S_2$, where option (3) is added to the offer set. This experiment exhibits a violation of the regularity assumption, 
since the probability of choosing option (2) increases from 50\% to 57\% when option (3) is added to the offer set. Hence, no model belonging to the RUM class can perfectly fit this dataset.}
\end{exmp}

\begin{table}[!h]
    \centering
    \begin{tabular}{lr@{\extracolsep{1cm}}r@{\extracolsep{0cm}}r}
    \toprule
     & & \multicolumn{2}{c}{Market share}\\
     \cmidrule{3-4}
     Model & Price (\$)  & $S_1$ & $S_2$ \\
     \midrule
     (1) Minolta X-370   &  $170$ & $.50$  & $.22$\\
     (2) Minolta 3000i   &  $240$ & $.50$  & $.57$\\
     (3) Minolta 7000i   &  $470$ & -      &   $.21$\\
     \bottomrule
    \end{tabular}
    \caption{Market share of three camera models in choice scenarios $S_1$, where respondents must choose between alternatives $\{1,2\}$, and $S_2$, where option $(3)$ is added to the offer set}
    \label{tab:MarketShare}
\end{table}

\revcs{The choice phenomena reported in Table \ref{tab:MarketShare} is an example of the so-called \textit{compromise effect}, where middle (i.e., compromise) options in terms of price and quality are preferred to extreme ones.} \revcs{ One may be tempted to handcraft an utility function based on price and quality features in order to explain the observed choice outcomes. However, this cannot be done without considering the \textit{assortment-dependent} effects on the attractiveness of products, which is inconsistent with the theory of rational choice.}

\rev{
 Violations of the regularity assumption may be induced by other cognitive biases as well. For example, in the context of grocery shopping, when two complementary products (e.g., pasta and tomato sauce) are present in the assortment, the perceived attractiveness of both is likely to increase. One may also observe asymmetric, or \textit{decoy} effects \citep[see, e.g.,][]{decoy_puto} when the addition of an option (the decoy) to the offer set increases the choice probability of another alternative perceived as better. This choice phenomena was popularized a choice experiment reported in \cite{ariely2008predictably}, in which a group of students was asked to choose among three possible subscription plans to ``The Economist'' magazine. For the sake of brevity, we report this experiment in Appendix \ref{ap:econ}, where we also show how the GSP model (see Section \ref{sec:model}) can explain the related choice outcomes. We refer the interested reader to \cite{gsp-berbeglia2018} for more examples on the topic.
}

 Such observations  motivated a recent interest in more general choice models, capable of overcoming the limitations of the RUM framework. Unfortunately, many of these choice models lack efficient estimation schemes, and their performance on non-RUM instances has not been well understood yet \citep[see, e.g.,][]{LoR-jagabathula}. Also, the minimal assumptions these models make about the distribution of choice probabilities may increase the risk of capturing spurious patterns from data, i.e., overfit.\footnote{\textcolor{black}{Mostly a concern in Statistics and Machine Learning, \textit{overfitting} refers to the situation when a model is too tailored to a specific data set (typically, the training data), and as such fails to generalize well to other data sets (e.g., the test data). Overfitting typically occurs either when the model is too general, or when the training data is not sufficiently representative for the ground truth. As a consequence, an overfit model may not yield accurate predictions on other data sets.}} \rev{This may be observed in the form of a model experiencing high variance in predictive accuracy when estimated on little amount of training data.} 
 Finding the delicate balance between flexibility and predictive accuracy is therefore of crucial importance for the practical utility of such models. The \textit{Generalized Stochastic Preference} (GSP) choice model, an extension of rank-based choice models introduced by \cite{gsp-berbeglia2018} to capture halo effects, is one of the recently proposed models that fits into this stream of literature. 
Despite being theoretically attractive, the estimation of the GSP choice model poses significant challenges both from the computational and predictive points of view. The authors suggest that estimation procedures originally developed for rational rank-based choice models \citep[see, e.g.,][]{farias13, Vulcano-marketDiscovery, bertsimas16-local_search} may be adapted to their irrational choice model. Nevertheless, no empirical study has been reported in order to assess the estimation efficiency and predictive accuracy of the GSP choice model.

\rev{Next to computational challenges to estimate such model, its flexibility also comes at the price of an increased risk of overfitting.}
%
\revcs{Even in the case of rational behaviors, it is known \citep[see, e.g., ][]{comparativeBerbeglia,jena2020partially} that the  estimation of general RUM models on little amounts of transactions data tends to be prone to overfitting. In the same line, on real-world data, \cite{HaloMNL} came to a similar conclusion. When estimated on small amounts of training data, their model, extending the MNL model to allow for pairwise interactions among products, did not outperform a simpler MNL. 
Hence, particular care should be taken to such issue for effectively estimating the GSP choice model, which can account for even higher orders of interactions among products.}

\vspace{0.1cm}
\paragraph{Contributions.}
%

%
In this work, we propose an estimation method for the GSP choice model. Specifically, we show how to use partially-ranked preferences to model irrational customer behaviors, and   how to efficiently estimate them from choice data by adapting the column generation approach proposed by \cite{jena2020partially}. Partially-ranked preferences allow us to circumvent several difficulties regarding the adaption of estimation methods for strictly ranked preferences. In particular, our objective is to train the choice model so as to maximize its predictive accuracy. This is different from \cite{farias13}, who focus on worst-case revenue prediction for a given assortment of items. Also, our estimation method can easily handle both rational and irrational customer behaviors. In contrast, it is not clear how the Mixed Integer Programming (MIP) formulation of the Market Discovery subproblem from \cite{Vulcano-marketDiscovery} should be adapted to allow for the discovery of irrational preferences. 
\textcolor{black}{Finally, the \textit{Growing Preference Tree} (GPT) algorithm of \cite{jena2020partially} 
provides a strong computational advantage in terms of scalability, especially important when dealing with irrational customer behaviors (discussed in the following) and generalizes well to unseen offer sets when tested on RUM instances.} The application of partially-ranked preferences for tackling the estimation of generalized stochastic preferences thus looks promising. An appealing property of our approach stems from the fact that the irrationality, and thus the flexibility of the choice model is increased in an adaptive, data-driven way. By increasing the set of possible customer behaviors only when required to better explain the given data, we may limit the risk of overfitting and speed up the estimation procedure. \revcs{To further reduce the risk of capturing spurious, high order interactions among products, we propose a new dominance rule among entering columns, prioritizing customer types with a small number of strictly ranked products and large indifference sets.}

We run an extensive set of experiments to assess the predictive performance of the proposed choice model. Using the methodology delineated by \cite{LoR-jagabathula},  we characterize the \textit{rationality loss} of both generated and real instances. This allows us to observe that irrational customer types can significantly improve predictive accuracy on instances presenting halo effects among alternatives. \revcs{We further show that our new criteria for discovering customer types can provide a further boost in predictive accuracy. Notably, our algorithm outperforms, on average, two baselines from the literature on irrational choice modeling, the Halo-MNL \citep{HaloMNL} and the Pairwise Choice Markov Chain (PCMC) \citep{PCMC}, when tested on real-world data.
}


\vspace{0.1cm}

\paragraph{Organization of the paper.}
In Section \ref{sec:lit_rev}, we review the literature on irrational choice models. In Section \ref{sec:model}, we introduce the GSP choice model from \cite{gsp-berbeglia2018} and our corresponding partially-ranked representation. We show how to estimate the proposed choice model in Section \ref{sec:estim}. The numerical results of our experiments on both synthetic and real instances are reported in Section \ref{sec:exps}. Finally, concluding remarks are reported in Section \ref{sec:conclusion}.

\section{Related work}
\label{sec:lit_rev}
\revcs{Our work spans several areas of research. In the following, we first review the literature from Psychology and Marketing, where several descriptive models, with little applicability from the predictive point of view, have been proposed to overcome the limitations of the RUM framework. We then survey the works from the Machine Learning and Operation Management communities, where discrete choice models of various levels of generality have been proposed.}

\vspace{0.5cm}

\paragraph{Descriptive theories of choice. }
In order to define the notion of a \textit{rational} agent, most economists rely on a set of consistency principles of rationality, which includes, among others, the aforementioned Regularity assumption and the more famous axiom of Independence of Irrelavant alternatives (IIA). This set of assumptions aims at describing how a rational agent is supposed to make her decisions across different offer sets. However, a vast body of literature has provided strong empirical evidence of choice behaviors incompatible with the theory of rational choice (we refer to \cite{Rieskamp2006} for an excellent overview on the topic). The RUM framework is flexible enough to explain most of these choice behaviors, but cannot account for violations of the Regularity assumption. 
To overcome such limitation, more general theories of choices have been developed in psychology, such as Decision Field theory \citep{Busemeyer1993, Roe2001} and the Leaky competing accumulator model  \citep{Usher2004a}. These models belong to the broader class of Sequential Sampling models, which mimic the evolution of the decision-making process over time, and can account for violations of the rationality principles, including the Regularity one. They lack, however, practical estimation algorithms, and are usually adopted from a descriptive point of view more than a predictive one. Other works, such as \cite{Tversky1993} and \cite{Rooderkerk2011}, embed alternatives into an attribute space, where context-dependent features are computed in order to determine the utility of each of the alternatives. These approaches have usually been applied to small, controlled experiments, and rely on the existence of two metric features, along which customer preferences are supposed to monotonically increase or decrease. This is a key difference with respect to our approach, where no item feature is supposed to be given. 

\vspace{0.2cm}

\paragraph{Discrete choice models escaping RUM. } Decomposing the utility into two components, item-specific and context-dependent, is also the starting point of \cite{HaloMNL} and \cite{CDM}, who propose a second-order extension of the MNL model in order to capture pairwise product interactions. However, these models do not subsume the RUM framework and thus, as pointed out by \cite{LoR-jagabathula}, are  not guaranteed to provide a better fit than RUM methods, even when applied to irrational instances. The same limitation holds for other models such as the General Attraction Model from \cite{GAM}, 
the Perception-adjusted choice model \citep{percp-adjusted-luce} and the General Luce Model \citep{echenique2015general}. \cite{Feng2017} propose a welfare-based framework, which subsumes the RUM framework and can be used to obtain choice models able to capture violations of the regularity assumption. The estimation of these choice models, however, is left by the authors as an open research question. Another general approach for which no empirical result has been reported is the Generalized Stochastic Preference choice model \citep{gsp-berbeglia2018}, an extension of rank-based choice models \citep[see, e.g.,][]{farias13, Vulcano-marketDiscovery} that allows for irrational customer behaviors. This model subsumes the RUM family of models and generalizes the non-RUM approach from \cite{KleinbergMU17} by allowing for heterogeneity in customer preferences. Despite its flexibility, the GSP choice model imposes some structure on the choice probabilities, and some examples are provided by the authors describing choice behaviors that do not belong to the GSP class. \cite{PCMC} propose the Pairwise Choice Markov Chain model, where each alternative is represented as a node of a continuos time Markov Chain. Given an offer set, the choice probabilities are given by the stationary distribution of the sub-chain consisting of the nodes indexed by the available alternatives. Although the PCMC choice model is able to capture both substitution and halo effects, it obeys the axiom of uniform expansion introduced by \cite{yellott1977uniformExpansion}. The authors argue that such property may be desirable in the context of discrete choice modeling.

\vspace{0.2cm}

\paragraph{Universal discrete choice models. }
Some more general choice models have been proposed in the literature, which are able to represent \textit{any} discrete choice function. In particular, \cite{osogami-rbm} propose an extension of the MNL model aming at capturing high-order product interactions. They show that the resulting model can be represented as a Restricted Boltzman Machine (RBM), a probabilistic graphical model whose units are divided into two groups, visible and hidden. Visible units are used to encode a binary representation of the offer set and of a given choice, while hidden units learn a latent representation of the input. Given enough hidden units, these models can represent any sort of irrational behavior. An approach based on tree ensembles has recently been proposed by both \cite{DecisionForest-gallego} and \cite{DecisionForest-misic}, who show that any discrete choice model can be represented as a distribution over decision trees. 

As previously mentioned, choice models with rather flexible structures pose some crucial challenges, whose solution greatly impacts the predictive accuracy of the trained choice models. In particular, one needs to balance between flexibility of the choice model, tractability of its estimation procedure, and risk of overfitting when limited amount of data is available. \rev{In this regard, it should be noted that our estimation procedure is non-parametric. Compared to the approach from \cite{osogami-rbm}, our model can adaptively increase the number of parameters in a data-driven way, as more data becomes available. This may help avoiding overfitting issues when only limited amounts of data, since a a model with a relatively small number of parameters will be learned in this case. The results from \cite{comparativeBerbeglia} seems to confirm this claim, showing Rank-Based choice models are relatively data-efficient, offering good generalization even with limited amounts of transactions data.  Also, our algorithm allows to include \textit{both} rational and irrational behaviors in the estimation process, thus providing a general method, well suited to a wide range of real case studies. The same is not true for other irrational models such as \cite{osogami-rbm}, which should be used with care on datasets where it is reasonable to assume that customers do act rationally.}  

\cite{DecisionForest-misic} and \cite{DecisionForest-gallego} tackle both the computational and generalization aspects by proposing regularization methods whose effect is to restrict the search space in a principled way. It may be argued, nevertheless, that less general choice models may be more effective in exploring search spaces that are smaller by definition, and that imposing some structure on the choice probabilities may provide important inductive bias to improve generalization over unseen offer sets when limited amount of data is available. This observation motivates the focus of this paper. In particular, we propose an estimation method for the Generalized Stochastic Preference choice model, which is flexible enough to subsume the RUM family of models and to capture halo effects, but still imposes some structure on the choice probabilities. 

 We conclude this section by mentioning an interesting line of work from the machine learning community, proposing general approaches based on Neural Networks to approximate the complex, high-order interactions among alternatives \citep[see, e.g.,][]{FATE, rosenfeld2019predicting, PointerNet}. Despite their flexibility, however, these models have only been applied to settings with product features and large number of training offer sets. Their adaptation to a setting close to ours, where no item featurization is given and the amount of offer sets seen at training time is relatively small, has not been explored yet and does not seem trivial.

\section{The choice model}
\label{sec:model}
\textcolor{black}{In this section, we review the Generalized Stochastic Preference model \citep{gsp-berbeglia2018} and extend it by allowing for partial ordering and indifference sets.}
Consider a set of products $\mathcal{N}=\{0,...,N-1\}$, with label 0 representing the \textit{no-purchase} option. 
\revcs{With a slight abuse of notation}, let then $\sigma$ denote both a subset of products in $\mathcal{N}$ and a linear order defined over such products, so that the rank (or position) of product $j$ according to $\sigma$ is given by $\sigma(j) \geq 1$.
%
\vspace{0.1cm}
\begin{definition}
\label{def:fullyrankedgsp}
A \textit{Generalized Stochastic Preference} \citep{gsp-berbeglia2018} $C(\sigma, i)$, consists of a ranking of products $\sigma \subseteq \mathcal{N}$, and an index $i$, with $1 \leq i \leq |\sigma|$. 
\end{definition}
\vspace{0.1cm}

Specifically, when faced with an offer set $S \subseteq \mathcal{N}$, \rev{a customer of type $k$, also referred to as $C_k(\sigma_k,i_k)$}, picks the alternative ranked $i^{th}$ in its subsequence $\sigma_k$ that only contains items also available in $S$. \rev{Further, we define $\sigma_{k,S} \subseteq \sigma$ as} the sequence of products obtained by removing from \rev{$\sigma_k$} every product $j \notin S$. The customer will then choose product $j^*$ so that $\sigma_{k,S}(j^*)=i$. If $|\sigma_{k,S}| < i$, the customer will leave without any purchase. The particular case of $i=1$ corresponds to customers who always pick their favorite (i.e., highest ranked) product among the available ones. For this reason, we refer to customers $C(\sigma, 1)$ as \textit{rational} behaviors, and to the index $i$ of a generalized stochastic preference as its \textit{irrationality level}. The GSP choice model is then defined by a probability distribution $\pmb{\lambda} \in \mathbb{R}^K$ over  $K$ customer types $\{C_k(\sigma_k,i_k)\}_{k=1}^K$, \rev{so that the probability of choosing item $j$ from assortment $S$ is given by

\begin{equation}
    P(j|S) = \sum_{k=1}^K \lambda_k \mathbbm{1}\{\sigma_{k,S}(j) = i_k\}.
\end{equation}
}

It should be noticed that, since every RUM choice model can be equivalently represented as a distribution over \textit{rational} stochastic preferences \citep[see, e.g.,][]{block1959random}, the GSP choice model naturally subsumes the RUM family of models. 

\revcs{\cite{KleinbergMU17} theoretically justify the use of ``irrational'', rank-based behaviors\footnote{In the work of \cite{KleinbergMU17}, the authors refer to \textit{position-selecting choice functions}, whose definition is essentially equivalent to the one of (irrational) customer behaviors from \cite{gsp-berbeglia2018}.} in order to capture compromise effects. In particular, the authors assume alternatives can be mapped to a one-dimensional embedding (i.e., utility) representing the alternatives overall evaluation by the decision-maker. Such embeddings can be given by their price or by a possibly complex function of their features. Alternatives can then be ranked in this embedding space according to their evaluation. Then, choosing the best ``compromise'' corresponds to selecting the option with rank $2 \leq i \leq  |S|-1 $, where $S$ is the set of available alternatives. To better illustrate the practical implications of this argument, Table \ref{tab:GSP_mod} exemplifies \citep{gsp-berbeglia2018} how the GSP model can explain the results of the experiment from \cite{simonson1992choice_exp} in Example \ref{example}.}

\begin{table}[h!]
      \centering
        \begin{tabular}{ccc@{\extracolsep{0.7cm}}r@{\extracolsep{1cm}}c@{\extracolsep{0.5cm}}c}
        \toprule
            \multicolumn{3}{c}{Customer Type} &  Probability & $S_1=\{1,2\}$ & $S_2=\{1,2,3\}$ \\
        \cmidrule{1-3}
            $k$   & $\sigma_k$ & $i_k$ & $\lambda_k$ & $\sigma_{k,S_1}$  & $\sigma_{k,S_2}$\\
        \midrule
            $1$   &  $(1,3,2)$   &  1   & 0.22 & $(\pmb{1},2)$ & $(\pmb{1},3,2)$\\
            $2$   &  $(2,3,1)$   &  1   & 0.29 & $(\pmb{2},1)$ & $(\pmb{2},3,1)$\\
            $3$   &  $(3,2,1)$   &  1   & 0.21 & $(\pmb{2},1)$ & $(\pmb{3},2,1)$\\
            $4$   &  $(3,2,1)$   &  2   & 0.28 & $(2,\pmb{1})$ & $(3,\pmb{2},1)$\\
        \bottomrule
        \end{tabular}
        \caption{GSP model from \cite{gsp-berbeglia2018} explaining the choice outcomes of Example \ref{example}. For each $\sigma_{k,S}$, we highlight in bold the chosen item $j: \sigma_{k,S}(j) = i_k$.} 
        \label{tab:GSP_mod}
\end{table}%


\begin{table}[!h]
    \centering
    \begin{tabular}{lr@{\extracolsep{1cm}}rr}
    \toprule
     &  & \multicolumn{2}{c}{Predicted Share} \\
     \cmidrule{3-4} 
     Model & Price (\$)  & $S_1$ & $S_2$ \\
     \midrule
     (1) Minolta X-370   &  $170$ & $\lambda_1 + \lambda_4 = .50$ & $\lambda_1 = .22$\\
     (2) Minolta 3000i   &  $240$ & $\lambda_2 + \lambda_3 = .50$ & $\lambda_2 + \lambda_4 = .57$\\
     (3) Minolta 7000i   &  $470$ & $-$ & $\lambda_3=.21$ \\
     \bottomrule
    \end{tabular}
    \caption{Predicted shares of three camera models in choice scenarios $S_1$, where respondents must choose between alternatives $\{1,2\}$, and $S_2$, where option $(3)$ is added to the offer set}
    \label{tab:PredShare}
\end{table}

\revcs{When only products $\{1,2\}$ are offered, customer $k=4$ chooses the cheapest one, i.e., option $(1)$. According to the rational theory of choice, this would imply that option $(1)$ has to be preferred to option $(2)$ independently of the other products in the offer set. This, however, would be in contradiction with the data at hand. Irrational choice behaviors, on the other hand, can account for assortment-dependent effects. By introducing option $(3)$ in the assortment, the same customer ends up choosing option $(2)$, thus implying that $(2)$ is preferred to $(1)$ in this case.} \revcs{ We refer to Appendix \ref{ap:econ} and \cite{gsp-berbeglia2018} for more examples showing how the GSP choice model can reproduce several experiments from the literature on behavioral economics showing evidence of irrational choice behaviors.}

From a modeling perspective, we note that including the 0 (i.e., no-purchase) option among the ranked alternatives has a useful implication in practice. In particular, contrary to the original formulation in \cite{gsp-berbeglia2018}, this allows us to capture violations of the regularity assumption also for the no-purchase option (we refer to Appendix \ref{ap:regularity0} for more details). Several studies, indeed, have shown that customers' willingness to purchase and overall satisfaction may decrease in the presence of too many alternatives among which a choice has to be made \citep[see, e.g.,][]{Iyengar2000, ParadoxOfChoice}.




The estimation of the GSP choice model poses significant computational challenges, given that the space of rational customer types alone is factorially large. Estimation procedures developed for rational, rank-based models such as those from \cite{Vulcano-marketDiscovery} and \cite{bertsimas16-local_search} cannot be easily adapted to account for learning irrational preferences, nor does their scalability look promising to tackle the even bigger search space implied by the presence of irrational behaviors \cite[see, e.g.,][]{comparativeBerbeglia, jena2020partially}. For these reasons, we decided to adopt the framework if partially-ranked preference sequences from \cite{jena2020partially} to represent generalized stochastic preferences. Besides providing a more intuitive, behavioral representation of an agent's decision process, partially-ranked preferences allow for fast estimation schemes and have been shown to generalize well on unseen offer sets. Starting from the observation that, for rational customer behaviors, low-ranked alternatives have a relatively low impact in explaining choice data, \cite{jena2020partially} propose to strictly rank only few, relevant alternatives for each preference list, while allowing for ties among the rest of them. Alternatives with the same rank may then be grouped into so-called \textit{indifference sets}. We thus provide the following definition:

\vspace{0.1cm}
\begin{definition}
\label{def2}
A \textit{partially-ranked preference with irrationality} $C(P(\sigma), I(\sigma), i)$, is defined by two sets of products  $P(\sigma)\subseteq \mathcal{N}$ and $I(\sigma) \subseteq \mathcal{N} \setminus P(\sigma)$, respectively, and a linear ordering $\sigma$ over the set of of alternatives $P(\sigma) \cup I(\sigma)$, so that $\sigma(j) = \sigma(j') \text{ for all } j,j' \in I(\sigma)$, and $\sigma(j) < \sigma(j')$ for all $j \in P(\sigma),j'\in I(\sigma)$.
\end{definition}
\vspace{0.1cm}

\rev{It follows from Definition \ref{def2} that a certain customer has no particular preference for products in $I(\sigma)$, which all have the same rank. Hence, we refer to $I(\sigma)$ as the \textit{indifference set} of that customer type.}
\rev{
Note, also, that the irrationality level of a partially-ranked preference is limited by $i \leq |P(\sigma)| + 1$, since alternatives in the indifference sets all have the same rank.
}
%
%
For ease of notation, let $P_S(\sigma) = P(\sigma) \cap S$ and $I_S(\sigma) = I(\sigma) \cap S$ denote the strictly ranked preference list and the indifference set, respectively, obtained after removing from $\sigma$ every product not available in a given offer set $S$.  A customer $C(P(\sigma), I(\sigma), i)$ will then pick the alternative ranked $i^{th}$ in $P_S(\sigma)$ if $i \leq |P_S(\sigma)|$, or an alternative chosen uniformly at random in $I_S(\sigma)$ when $ |P_S(\sigma)| < i \leq |P_S(\sigma) \cup I_S(\sigma)|$. When $i > |P_S(\sigma) \cup I_S(\sigma) |$, the customer leaves without any purchase.  
\begin{table}[!htbp]
    \centering
    \begin{tabular}{crrrrr}
        \toprule
         \multicolumn{2}{c}{offer set} & $P_S(\sigma)$ & $I_S(\sigma)$ & $\text{Choice}_{C_1}$ & $\text{Choice}_{C_2}$\\
         \bottomrule
         $S_1$ & $\{2,5,1\}$&   (2,5) & $\{1\}$ & 2& 5\\
          $S_2$ & $\{2,1,4\}$& (2) & $\{1,4\}$ & 2 & $\sim \text{Unif}\{1,4\}$ \\
         $S_3$ & $\{1\}$ &    $()$ & \{1\} & 1& 0\\
          $S_4$ & $\{1,4\}$ & $()$ & $\{1,4\}$ & $\sim \text{Unif}\{1,4\}$ & $\sim \text{Unif}\{1,4\}$\\
         \bottomrule
        \end{tabular}
    \caption{Example: choice behavior of two customers $C_1\big((2,3,5),\{1,4\},1\big)$ and $C_2\big((2,3,5),\{1,4\},2\big)$ across different offer sets.}
    \label{tab:EXAMPLE}
\end{table}

\rev{
While Section \ref{sec:estim} elaborates on how to estimate such preference sequences from transaction data, 
}
Table \ref{tab:EXAMPLE} gives an example of the choice behavior of two hypothetical customers $C_1\big((2,3,5),\{1,4\},1\big)$ and $C_2\big((2,3,5),\{1,4\},2\big)$, who differ only by their irrationality level. As a consequence, for each offer set $S$, we have that $P_S(\sigma_1) = P_S(\sigma_2)$ and $I_S(\sigma_1) = I_S(\sigma_2)$. \revcs{Specifically, the first row of Table \ref{tab:EXAMPLE} corresponds to the case where $i_1=1 \leq 2 = |P_S(\sigma_1)|$ and $i_2=2 \leq 2 = |P_S(\sigma_2)|$. The two customers will then select the items $j_1=2$ and $j_2=5$ with ranks 1 and 2, respectively, in $|P_S(\sigma)|$. For all the other assortments, however, we have that $i_2 = 2 > 1 \geq |P_S(\sigma)|$, hence Customer 2 will either pick an item uniformly at random from the indifference set (assortments $S_2$ and $S_4$) or leave without any purchase (assortment $S_3$). The same reasoning can be applied to obtain the choice of Customer 1 in the remaining assortments.} 
%


\section{Estimation procedure}
\label{sec:estim}
\cite{jena2020partially} have shown that \textit{rational}, partially-ranked preferences can be efficiently learned from data, using a \textit{ Growing Preference Tree} (GPT) algorithm. 
\rev{In this section, we first review the GPT  estimation framework, and then show how to extend the algorithm to additionally handle partially-ranked preferences with irrationality.}

We assume that training data is available in the form of $T$ observations $\mathcal{T}=\{(S_t, c_t)\}_{t=1}^T$ with $S_t$ and $c_t$ representing the offer set and the choice, respectively, that have been observed in period $t$. Let $\mathcal{S}_{train}=\{S_1,...,S_M\}$  denote the collection of $M$ offer sets over which choice data is available. We can further preprocess dataset $\mathcal{T}$ in order to obtain a vector of empirical probabilities $\pmb{v} \in \mathbb{R}^{N \cdot M}$ so that, for each $j \in \mathcal{N}$ and $S \in \mathcal{S}_{train}$, the probability of item $j$ being chosen from offer set $S$ is given by 
\begin{equation*}
    v_{j,S} = \frac{\sum_{t=1}^T \mathbbm{1}\{S_t=S, i_t=j\}}{\sum_{t=1}^T \mathbbm{1}\{S_t=S\}}.
\end{equation*}

\subsection{Non-parametric estimation framework}

The GPT algorithm fits into the  general column-generation framework proposed for the estimation of a general class of nonparametric choice models by \cite{Vulcano-marketDiscovery}.
In line with this framework, customer behaviors are represented as a choice matrix $\pmb{A} \in \mathbb{R}^{(N\cdot M) \times K}$, encoding $K$ behaviors for $M$ offer sets, whose elements give the probability of customers choosing an item from a given offer set. In particular, based on the choice behaviors of a partially-ranked list with irrationality $i$ defined in Section \ref{sec:model}, the elements of the matrix $\pmb{A}$ may be computed as follows:

\begin{equation}
A^k_{j,m} = 
  \left\{
      \begin{aligned}
            & 1 & \text{if } j \in P_{S_m}(\sigma) \text{ and } j \text{ ranked } i^{th} \text{ in } P_{S_m}(\sigma),\\
            & \dfrac{1}{|I_{S_m}(\sigma)|} &  \text{if } j \in I_{S_m}(\sigma) \text{ and } |P_{S_m}(\sigma)| < i \leq |P_{S_m}(\sigma) \cup I_{S_m}(\sigma)|   \\
            & 0  & \text{otherwise}. \\
      \end{aligned}
    \right.
\label{eq:computeA}
\end{equation}
 Given a distribution $\pmb{\lambda} \in R^K$ over the customer types, the predicted probability $x_{j,m}$ of a random customer choosing  alternative $j$ from the offer set $S_m$ is then given by $x_{j,m} = \sum_k A^k_{j,m} \lambda_k$. One can thus define the \textit{best} distribution $\pmb{\lambda}$, that is, the one for which the predicted probabilities are the closest to the observed ones, and obtain $\pmb{\lambda}$ by solving the following optimization problem: 
%
%
\begin{subequations}
\label{f:F1-all}
\begin{alignat}{2}
\quad  \min_{ \pmb{\lambda}, \pmb{x} }  \ \
& \mathcal{L}(\pmb{x},\pmb{v})
\label{f:F1OF}
\\
\text{s.t.} \ \
& \pmb{A} \pmb{\lambda} = \pmb{x}  
\label{f:F1C1}
\\
& \pmb{1}^T \pmb{\lambda} = 1 
\label{f:F1C2}
\\
& \pmb{\lambda}  \geq 0.
\label{f:F1C3}
\end{alignat}
\end{subequations}

Here, $\mathcal{L}(\pmb{x}, \pmb{v})$ can be any convex loss function measuring the distance between the predicted probabilities $\pmb{x}$ and the observed ones $\pmb{v}$. For example, one may minimize the $L_1$ error between the two probability distributions, in which case we have

\begin{equation}
    \label{eq:l1}
    \mathcal{L(\pmb{x}, \pmb{v}}) = \sum_{S \in \mathcal{S}_{train}}\sum_{i \in S}|x_{i,S} - v_{i,S}| .
\end{equation}

Minimizing the $L_1$ error generally leads to sparse models. Further, objective function (\ref{eq:l1}) can be easily linearized \citep[see, e.g.,][]{bertsimas16-local_search} and is therefore computationally amenable.

Another popular measure of the distance between two probability distributions is the Kullback-Leibler. It is strictly convex and leads to the same solution as Maximum Likelihood Estimation \citep[see, e.g.,][]{LoR-jagabathula}.  It is computed as
\begin{equation}
    \label{eq:kl}
    \mathcal{L}(\pmb{x},\pmb{v}) = -\frac{1}{T} \sum_{S \in \mathcal{S}_{train}} T_S \sum_{i \in S}v_{i,s} ~ log \frac{x_{i,S}}{v_{i,S}}\text{,}
\end{equation}

\noindent
where $T_S$ is the number of samples showing $S$ as offer set.

\subsection{Discovering new rational and irrational customer types}
\label{sec:newcusttypes}


\rev{Solving problem (\ref{f:F1-all}) over the factorially large set of all possible customer types is not tractable. Hence, the master problem (\ref{f:F1-all}) is first initialized with a restricted set of customer behaviors and is solved to find the corresponding probability distribution $\pmb{\lambda}$ that best fits the training data. 
In each further iteration, new relevant behaviors are discovered by solving a subproblem and added to the master problem, which then adjusts the probability distribution $\pmb{\lambda}$ over the new set of behaviors.
%
The algorithm terminates once a predefined stopping criteria is met.}

In particular, let $\pmb{\alpha} \in \mathbb{R}^{N\cdot M}$ and $\nu \in \mathbb{R}$ denote the dual variables associated with constraints (\ref{f:F1C1}) and (\ref{f:F1C2}), respectively. The customers (i.e., preference sequences) worth adding to the model in order to improve its fit of the data are those whose corresponding choice vector $\pmb{a}$ \revcs{(column of A)}, computed as in (\ref{eq:computeA}), has a \revcs{negative (reduced) cost $c(\sigma)= -\pmb{\alpha}^T\pmb{a} - \nu$}. While finding new preference sequences with \revcs{negative costs} generally tends to be computationally expensive, the GPT algorithm exploits the structure of partially-ranked preferences to efficiently identify such columns. Indeed, partially-ranked preferences allow for using an efficient tree-like data structure, where deeper levels correspond to behaviors with more refined ranked lists. 
\revcs{Specifically, any sequence of products obtained from such a tree by traversing the path from the root to a given node, corresponds to the preference sequence $P(\sigma_k)$ of a certain customer $C_k$. It then follows that a behavior $C_j$ is considered a sub-behavior of $C_k$, if $P(\sigma_j) = (P(\sigma_k),\ell)$ with $\ell \in I(\sigma_k)$. When searching for new customer behaviors, one may thus restrict the search for relevant customer types among the sub-behaviors of $\{C_1,...,C_K\}$, at any given iteration.}

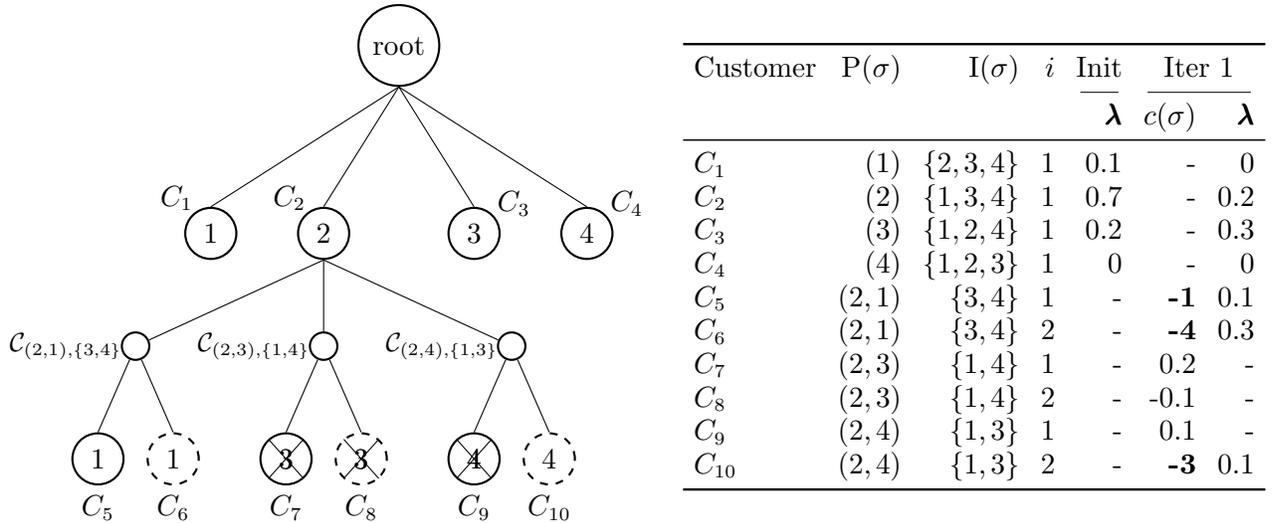
\begin{figure}[!htpb]
    \begin{minipage}{0.5\linewidth}
		\centering
	\begin{tikzpicture}[
		roundnode/.style={circle, draw=black, thick, minimum size=2mm},
        ]

        \node[roundnode] at (2.5,5.5)        (root)        {root};
        \node[roundnode, label={[label distance=-1.5mm]120:$C_1$}] at (0,3)       (single1)        {1};
        \node[roundnode, label={[label distance=-1.5mm]120:$C_2$}] at (1.5,3)          (single2)        {2};
        \node[roundnode, label={[label distance=-1.5mm]30:$C_3$}] at (3.5,3)          (single3)        {3};
        \node[roundnode, label={[label distance=-1.5mm]30:$C_4$}] at (5,3)        (single4)        {4};
        
        \node[roundnode, label={[label distance=-1.5mm]left:\small{$\mathcal{C}_{(2,1),\{3,4\}}$}}] at (-1, 1.5) (middle_56) {};
         \node[roundnode, label={[label distance=-1.5mm]left:\small{$\mathcal{C}_{(2,3),\{1,4\}}$}}] at (1.5, 1.5) (middle_78) {};
        \node[roundnode,label={[label distance=-1.5mm]left:\small{$\mathcal{C}_{(2,4),\{1,3\}}$}}] at (4, 1.5) (middle_910) {};
        
        \node[roundnode, label=below:$C_5$] at (-1.5, 0) (1rat) {1};
        \node[roundnode, label=below:$C_6$, dashed] at (-0.5, 0) (1irrat) {1};
        
        \node[roundnode, label=below:$C_7$] at (1, 0) (3rat) {3};
        \node[cross out, draw] at (1, 0) (X3rat) {3};
        \node[roundnode, label=below:$C_8$, dashed] at (2, 0) (3irrat) {3};
        \node[cross out, draw] at (2, 0) (X3irrat) {3};
        
        \node[roundnode, label=below:$C_9$] at (3.5, 0) (4rat) {4};
        \node[cross out, draw] at (3.5, 0) (X4rat) {4};
        \node[roundnode, label=below:$C_{10}$, dashed] at (4.5, 0) (4irrat) {4};
        
        \draw[-] (root.south) -- (single1.north);
        \draw[-] (root.south) -- (single2.north);
        \draw[-] (root.south) -- (single3.north);
        \draw[-] (root.south) -- (single4.north);
        
        \draw[-] (single2.south) -- (middle_56) -- (1rat.north);
        \draw[-] (middle_56) -- (1irrat.north);
        \draw[-] (single2.south) -- (middle_78) -- (3rat.north);
        \draw[-] (middle_78) -- (3irrat.north);
        \draw[-] (single2.south) -- (middle_910) -- (4rat.north);
        \draw[-] (middle_910) -- (4irrat.north);


    \end{tikzpicture}
	\end{minipage}\hfill
	\begin{minipage}{0.45\linewidth}
		\centering
		\begin{tabular}{lrrrrrr}
			\toprule
			Customer        & P($\sigma$) & I($\sigma$) & $i$ &  \multicolumn{1}{c}{Init} & \multicolumn{2}{c}{Iter 1}\\
			\cmidrule(lr){5-5} \cmidrule(lr){6-7}
			& & & &  $\pmb{\lambda}$ & $c(\sigma)$ & $\pmb{\lambda}$ \\
			\midrule
			$C_1$ & $(1)$ & $\{2,3,4\}$ & 1 & $0.1$ &- & 0\\
			$C_2$ & $(2)$ & $\{1,3,4\}$ & 1 &  $0.7$ &- & 0.2\\
			$C_3$ & $(3)$ & $\{1,2,4\}$ & 1 &  $0.2$ &- & 0.3\\
			$C_4$ & $(4)$ & $\{1,2,3\}$ & 1 &  $0$   &- & $0$\\
			$C_5$ & $(2, 1)$ & $\{3,4\}$ & 1 &  - & \textbf{-1} &$0.1$\\
			$C_6$ & $(2, 1)$ & $\{3,4\}$ & 2 &  - & \textbf{-4} & $0.3$\\
			$C_7$  & $(2, 3)$ & $\{1,4\}$ & 1 &  - & 0.2 & - \\
			$C_8$  & $(2, 3)$ & $\{1,4\}$ & 2 &  - & -0.1 & - \\
			$C_9$  & $(2, 4)$ & $\{1,3\}$ & 1 &  - & 0.1&- \\
			$C_{10}$ & $(2, 4)$ & $\{1,3\}$ & 2 &  - &\textbf{ -3} & $0.1$\\
			\bottomrule
        \end{tabular}
	\end{minipage}
	\caption{(Left) Search-Tree of GPT for finding new behavior, on a toy example with four products, after two iterations. A path in the tree corresponds to a sequence of strictly ranked products. Dashed nodes correspond to irrational behaviors. (Right) The explicit behaviors description, with corresponding probabilities and costs at a given iteration.}
	\label{fig:behavior-split}
\end{figure}

\revcs{To better illustrate how the search-tree is gradually explored during the GPT procedure, we report in Figure \ref{fig:behavior-split} the tree resulting from the initialization step, followed by one iteration of the GPT algorithm on a toy example, where the universe of products consists of four alternatives, i.e., $\mathcal{N}=\{1,2,3,4\}$. 
The algorithm starts by initializing the tree with $N$ \textit{rational} customer types $C_1, \dots, C_N$ such that $P(\sigma_k)=k$ and $I(\sigma_k)=\mathcal{N} \setminus \{k\}$. We then solve the restricted master problem (\ref{f:F1-all}) over this initial set of behaviors, in order to obtain a first distribution $\pmb{\lambda}$ over the $N$ customer types, and the values of the dual variables $\pmb{\alpha}$ and $\nu$. After the initialization, the generation of new candidate behaviors to include in the master problem at each future iteration requires three steps:}

\revcs{
\begin{enumerate}

    \item \textbf{Sampling: } We select $\gamma$ behaviors from the existing nodes via random sampling according to probability distribution $\pmb{\lambda}$. In the example in Figure \ref{fig:behavior-split}, we have sampled $\gamma=1$ behavior in the first phase: namely, $C_2$ (i.e., $k=2$ with probability $\lambda_2=0.7$).
    
    \item \textbf{Sub-behavior generation: } For every sampled behavior, we first generate $|I(\sigma_k)|$ \textit{rational} sub-behaviors ($C_5$, $C_7$ and $C_9$ in the example). These new behaviors $(P(\sigma), I(\sigma),1)$  
    are obtained by removing one item from the indifference set of the parent node ($C_2$ in the example) and adding it to the end of its strictly ranked preference sequence ($P(\sigma_2)$). 
    Such rational sub-behaviors (with irrationality level $i=1$) correspond to those also added by \cite{jena2020partially}. We here extend this approach
    by additionally generating the \textit{irrational} counterparts with irrationality levels $i$ with $1 < i \leq |P(\sigma)|+1$ ($C_6$, $C_8$ and $C_{10}$ in the example).
    
    \item \textbf{Sub-behavior selection: } We then compute the reduced costs $c(\sigma)$ for each candidate sub-behavior and select the $\delta$ behaviors with the best (i.e., smallest) reduced costs, where $\delta$ is a pre-specified hyper-parameter (in the above example, $\delta=3$, selecting customers types $C_5,C_6$ and $C_{10}$). The nodes corresponding to the remaining customer types are pruned from the search-tree (nodes $C_7, C_8$ and $C_9$ in the example).
\end{enumerate}
}

\rev{Identifying relevant customer behaviors based solely on their reduced costs may not be robust in general when dealing with irrational customer behaviors. We elaborate more on the topic in Section \ref{sec:newrule} where, based on behavioral considerations, we propose a superior selection criteria for the identification of relevant customer types.}

Observe that customers that differ only in their irrationality level (such as $C_5$ and $C_6$ in our example) generate the same sets of sub-behaviors. 
\rev{Hence, only one of these nodes (either $C_5$ or $C_6$ in our example) has to be considered to evaluate and generate further sub-behaviours.} 
We therefore consider only one node among the set of behaviors $\mathcal{C}_{P(\sigma),I(\sigma)} = \{ C_k(P(\sigma_k), I(\sigma_k), i_k): P(\sigma_k) = P(\sigma) \text{ and } I(\sigma_k)=I(\sigma), k=1,...,K \}$, with probability $\tilde{\lambda}_{P(\sigma),I(\sigma)} = \sum_{k : C_k \in \mathcal{C}_{P(\sigma),I(\sigma)}} \lambda_k$.  \revcs{Considering the example in Figure \ref{fig:behavior-split}, at the second iteration the preference list $((2,3), \{1,4\})$ would then be sampled with probability $\tilde{\lambda}_{((2,3)\{1,4\})} = \lambda_5 + \lambda_6=0.4$.}

We highlight two major advantages of the exploration strategy employed by the GPT algorithm:
\begin{enumerate}
    \item The number of strictly ranked objects increases in an adaptive, data-driven way, with more refined preference lists added only when needed. Besides allowing to avoid the computational burden of strictly ranking all the products in a preference list, \cite{jena2020partially} show that the explanatory power of indifference sets tends to improve generalization on new offer sets.
    \item Since the irrationality level $i$ of a partially-ranked behavior is bounded by the number of strictly ranked products, i.e., $1\leq i \leq |P(\sigma)|$, the GPT search procedure prioritizes customers with low irrationality levels.  This seems to be a behaviorally-plausible inductive bias, which may reduce the risk of overfitting, especially when only a limited amount of data is available.
\end{enumerate}

\vspace{0.1cm}
\rev{
\paragraph{Computational complexity. }
As described above, each iteration of the GPT procedure involves sampling $\gamma$ behaviors $C_k(P(\sigma_k),I(\sigma_k),i_k)$ and, for each them, generating $|I(\sigma_k)| \cdot (|P(\sigma_k)|+1)$ sub-behaviors, i.e., one for each item $j \in I(\sigma_k)$ and irrationality level $i$, with $1 \leq i \leq |P(\sigma_k)|+1 $. The reduced cost $c(\sigma)$ of a sub-behaviour $\sigma$ can be obtained in $O(M)$ \citep{jena2020partially} from the reduced cost of its parent node. Therefore, computing the reduced costs of all rational and irrational candidate sub-behaviors at a given iteration has a complexity of $O(N^2M)$ (compared to $O(NM)$ for the rational approach). While this is the theortical worst-case complexity, our experiments in Appendix \ref{ap:learning_stats}  show that the GPT tends to produce relatively short preference lists $P(\sigma)$, ranking as few as 3 products, on average. Further, \cite{jena2020partially} showed that even on large (rational) instances with up to 1,000 products, the length of the produced strict preference lists tends to be similarly small, and rather independent from the number of products. The computational burden of each GPT iteration is therefore significantly mitigated in practice, since is tends to generate few irrational behaviors.
}


\subsection{A new dominance rule to select relevant customer types} 
\label{sec:newrule}

\revcs{
 In this section, we elaborate on how to identify new customer behaviors that improve the fit to the training data and are more likely to generalize well on test data. The \textit{de facto} scoring method in column generation, and therefore for evaluating the quality of candidate behaviors in the GPT estimation procedure, is exclusively based on their reduced costs $c(\sigma)$.
 This is based on the hypothesis that a parsimonious model
 tends to generalize well on unseen offer sets, which is widely adopted in the literature \citep[see, e.g.,][]{Vulcano-marketDiscovery,bertsimas16-local_search,jena2020partially}.}
\revcs{
However, when dealing with irrational behaviors, such criterion may lead to capturing an excessive number of spurious interactions among products, especially in the case of scarce data availability.

}

\revcs{
In the following, we argue that the use of customer behaviors with small numbers of strictly ranked products, and, as a consequence, larger indifference sets, may drastically reduce the risk of overfitting. 
Using rational behaviors only, this is naturally achieved by the design of the GPT algorithm, which starts by generating behaviors with small numbers of strictly ranked products. 
While, in the rational case, strictly ranking only a few products with respect to the total number of existing ones (e.g., 5 out of 100 products) may be sufficient to reduce the risk of overfitting, this may not be the case when considering irrational behaviors. In fact, here, the added risk of overfitting increases quickly with the number of possible interactions within the sequence of strictly ranked products.
}

\rev{
As an example, consider the case in which we aim to capture the positive relation of item $1$ on item $2$ in a total universe of $N=5$ products (for simplicity, we here do not use the no-purchase option in the preference sequences or the assortments), based on the observed choices of product 3 from assortment $S_1=\{2,3\}$ and product 2 from assortment $S_2=\{1,2,3\}$. When estimating the choice model, several irrational preference sequences may fit these transactions, for example, either $C\big((1,2),\{3,4,5\}), 2\big)$ or $C\big((1,2,3,4,5),\{\}, 2\big)$. The former (with few strictly ranked products) is clearly the more precise one, and is less prone to overfitting, while the latter is an example of a more general one, prone to a higher risk of overfitting. While the latter sequence seems to have unnecessarily many strictly ranked products, in practice, when dealing with limited transaction data, such sequences may be selected due to their lower reduced costs.
A choice model using such a sequence may, depending on the unseen offer sets, extrapolate up to $|P|\cdot(|P|-1)/2$ possible pairwise interactions, which are illustrated in Table \ref{tab:interactions}. Note that all except for one of these interactions are spurious and therefore overfit the training data. In contrast, the former sequence with only two strictly ranked products ``deactivates'' the influence of product 1 on any product other than 2.
\begin{table}[!htpb]
    \centering
    \begin{tabular}{crcrc}
        \toprule
         Positive interaction   &  $S_1$ & $\text{Choice}_{S_1}$ & $S_2$  & $\text{Choice}_{S_2}$\\
         \midrule
         $1 \rightarrow 2$      &  $\{2,3\}$ & 3 & $\{1,2,3\}$ & 2 \\
         $1 \rightarrow 3$      &  $\{3,4\}$ & 4 & $\{1,3,4\}$ & 3 \\
         $1 \rightarrow 4$      &  $\{4,5\}$ & 5 & $\{1,4,5\}$ & 4 \\
         $1 \rightarrow 5$      &  $\{5\}$ & 0 & $\{1,5\}$ & 5 \\
         $2 \rightarrow 3$      &  $\{3,4\}$ & 4 & $\{2,3,4\}$ & 3 \\
         $2 \rightarrow 4$      &  $\{4,5\}$ & 5 & $\{2,4,5\}$ & 4 \\
         $2 \rightarrow 5$      &  $\{5\}$ & 0 & $\{2,5\}$ & 5 \\
         $3 \rightarrow 4$      &  $\{4,5\}$ & 5 & $\{3,4,5\}$ & 4 \\
         $3 \rightarrow 5$      &  $\{5\}$ & 0 & $\{3,5\}$ & 5 \\
         $4 \rightarrow 5$      &  $\{5\}$ & 0 & $\{4,5\}$ & 5 \\
         \bottomrule
    \end{tabular}
    \caption{Some of the positive interactions $j \rightarrow j'$ implied by a single customer behavior $C\big((1,2,3,4,5), 2\big)$. In particular, by considering the offer set $S_2=S_1 \cup \{j\}$, customer's choice changes in favour of product $j'$.}
    \label{tab:interactions}
\end{table}
While it is, of course, not possible to know beforehand which products  really do interact which each other, it becomes immediate that, in an effort of reducing the risk of overfitting, it is desirable to prioritize behaviors with a small number of strictly ranked products.
}

\rev{
The number of spurious interactions to which an irrational customer behavior may lead to can be quantified exactly, as demonstrated in the following observation.
\begin{observation}
\label{proposition:NbSpuriousInteractions}
(\textbf{Number of spurious positive interactions}):
Let $C(P(\sigma), I(\sigma), i)$ be a partially-ranked preference sequence with strictly ranked products $P(\sigma)$, indifferent set $I(\sigma)$ and level of irrationality $i$. Depending on the unseen offer set, a choice-model based on $C$ may extrapolate up to $\sum\limits^{|P(\sigma)|-1}_{j=i-1} {j \choose i-1}$ positive interactions of degree $i$.
\end{observation}
\textbf{Proof: } see Appendix \ref{ap:proofObs1}.
}

\rev{
Observation \ref{proposition:NbSpuriousInteractions} implies that the number of spurious interactions, leading to potential overfitting, does not necessarily grow with a higher level of irrationality, but rather with the number of strictly ranked products.
We note that this observation is not limited to our definition of partially-ranked preference sequences, but generally applies to the case of behaviors defined by the Generalized Stochastic Preference model from \cite{gsp-berbeglia2018}.
}

\rev{
We now propose a new dominance rule to select the candidate behaviors to be included in the master problem (\ref{f:F1-all}). The rule aims at striking a delicate balance between the number of strictly ranked products and the reduced costs of a candidate behavior:
\begin{enumerate}
    \item For every behavior $\sigma$ belonging to the the set of candidate behaviors at a given iteration, we use equation (\ref{eq:computeA}) to compute its choice vector representation $\pmb{a}$, and its cost $c(\sigma) = -\pmb{\alpha}^T\pmb{a} - \nu$;
    \item We then lexicographically sort the candidate behaviors: first in increasing order of their number of strictly ranked products, i.e., $|P(\sigma)|$, and second in non-increasing order of their reduced costs $c(\sigma)$. This sorting operation induces a ranking $\pi$ among the candidate behaviors;
    \item Ranking $\pi$ itself does not guarantee that its highest-ranked candidates have negative reduced costs. While it is reasonable to add some candidates with non-negative costs (which have shown in experiments to be beneficial for future iterations), we have to ensure also adding candidates with negative costs in order to improve the model fit. We therefore start considering candidates in ranking $\pi$ at the highest rank corresponding to a behavior with negative cost:  \underbar{$\pi$}$=\argmin_{k}\{\pi(\sigma_k) | c(\sigma_k) < 0\}$. We then select all candidate behaviors $\{k | \text{\underbar{$\pi$}} \leq \pi(\sigma_k)\leq \text{\underbar{$\pi$}}+\delta-1\}$, for a pre-defined $\delta$.
\end{enumerate}

As will be shown in Section \ref{sec:exps}, the use of this selection rule improves the predictive performance of the irrational GPT algorithm, particularly on the real-world data set.
}

\begin{table}[!htbp]
    \centering
    \renewcommand{\arraystretch}{1.3} 
    \begin{tabular}{
    l
    >{\raggedleft\arraybackslash}p{1cm}
    >{\raggedleft\arraybackslash}p{1cm}
    >{\raggedleft\arraybackslash}p{1cm}
    >{\raggedleft\arraybackslash}p{1cm}
    >{\raggedleft\arraybackslash}p{1cm}
    >{\raggedleft\arraybackslash}p{1cm}
    >{\raggedleft\arraybackslash}p{1cm}}
    \toprule
        $k$  & 3 & 1 & 4& 5 & 2 & 6 & 7 \\ 
        \cmidrule{2-8}
         $|P(\sigma_k)|$  & 1 & 2 & 2 & 2 & 3 & 3 & 4\\
         $c(\sigma_k)$ & 0.2 & $-2$ & $-1$ & 0.01 & $-0.1$ & 3 & $-3$\\
         $\pi(\sigma_k)$  & 1 & 2 & 3& 4 & 5 & 6 & 7 \\
        \midrule
        Criterion  & \multicolumn{7}{c}{Entering Columns} \\
         \cmidrule(l){2-8}
        $c(\sigma_k)$ & & \checkmark & \checkmark & & & & \checkmark \\
        $(|P(\sigma_k)|, c(\sigma_k))$   & & \checkmark& \checkmark & \checkmark & & & \\
    \bottomrule
    \end{tabular}
    \caption{Difference in selected columns using a criterion based solely on the cost $c(\sigma_k)$, and one where columns are first ordered based on the number of strictly ranked products $|P(\sigma_k)|$}
    \label{tab:dominancerule}
\end{table}

\rev{Table \ref{tab:dominancerule} exemplifies the selection process according to each of the two possible selection criteria for seven candidate behaviors. The first four lines show, respectively, the behavior id, the number of strictly ranked products, the cost of each behavior and the resulting ranking $\pi$.
Using cost $c(\sigma)$ as selection critera would lead to selecting customers with smallest cost $c(\sigma)$, namely $C_1, C_4$ and $C_7$. However, notice that customer $C_7$ strictly ranks a relatively high number of products and, based on Observation \ref{proposition:NbSpuriousInteractions}, may imply a high number of spurious interactions among products. By using our proposed selection criterion, instead, we select behaviors with only two strictly ranked products. In particular, assuming $\delta=3$, we have \underbar{$\pi$}$=2$ and will select behaviors $C_1, C_4$ and $C_5$.}


\rev{
We conclude this section highlighting connections between our proposed dominance rule and \textit{consideration sets}, a well known concept in the Marketing literature \citep[see, e.g.,][]{hauser2010considreview}. This concept refers to the fact that consumers usually pay attention to only a small subset of all existing items (either due to a limited attention budget or to avoid the cognitive burden of searching through a possibly very large set of products), and will select an item from this subset. 
}

\rev{While the estimation of consideration sets is object of a separate branch of literature \citep[see, e.g.,][]{aouad2015assortment, jagabathula2019inferring}, our notion of strictly ranked sets approximate consideration sets, while still attributing a non-zero probability to products in the indifference set.  It has also been shown in the literature on brand choice that the size of consideration sets tends to be relatively small, usually consisting of 2 to 5 brands \citep[see, e.g.,][]{hauser1990evaluation}. This suggests, from a behavioral point of view, that prioritizing behaviors with a small number of strictly ranked products (and therefore lower orders of interactions among products) may provide good inductive bias for generalizing well on unseen offer sets.}

\section{Computational results}
\label{sec:exps}
In this section, we report the results of our experiments on both synthetic and real datasets. The goal is to understand whether irrational, partially-ranked behaviors can improve predictive accuracy on new offer sets. In all our experiments, we compare three variants of the partially-ranked choice model estimated using GPT, namely GPT-R, \rev{which corresponds to the approach of \cite{jena2020partially}, thus restricting the space of customer behaviors to the rational ones only, and its extensions GPT-I and GPT-IC, which include irrational behaviors in the estimation process. Of these, the former selects customer behaviors based solely on their costs, while the latter use the dominance rule proposed in this work, aiming at prioritizing customer behaviors with small \textit{consideration sets}, i.e., with a small number of strictly ranked products, and thus focusing  on sparse, low-order interactions among products}. We further compare the GPT-based approaches with three benchmarks:  the enumerative rank-based choice model (RB-R) with fully-ranked lists, obtained by enumerating all possible preferences (i.e., permutations over products)%
\footnote{Let $m$ denote the maximum number of products missing from any assortment. As noted in \cite{LoR-jagabathula} and \cite{Honhon2012}, only $O(N^m)$ permutations of $m+1$ products need to be generated, since products with rank greater than $m+1$ will never be chosen.  }
, the pairwise choice markov chain (PCMC) proposed by \cite{PCMC}, and the Halo-MNL choice model from \cite{HaloMNL}. Section \ref{subsec:synthetic_exps} focuses on the generalization performances of the various approaches on a set of synthetic instances. In Section \ref{subsec:real_exps}, \rev{we test the models on the IRI Academic dataset \citep{bodea2009data}, consisting of transaction data from a large grocery store chain}.  

  \paragraph{Estimation of the choice models.}
    Following \cite{LoR-jagabathula}, we train RB-R by minimizing the average Kullback-Leibler (KL) divergence (\ref{eq:kl}) between predicted probability distributions and the empirical ones over training offer sets. All the other approaches are trained by Maximum Likelihood Estimation. As already observed, it is well known that minimizing the KL divergence is equivalent to maximum likelihood estimation in terms of optimal solution retrieved \citep[see, e.g.,][]{LoR-jagabathula}. 
    For the GPT-based approaches, we stop the training procedure when the difference in the log-likelihood of the data at two consecutive iterations is statistically insignificant, as proposed by \cite{Vulcano-marketDiscovery}%
    \footnote{Let $\mathcal{L}^k$ and $\chi^2(\beta)$ denote the log-likelihood at iteration $k$ and the critical value of the chi-squared distribution with $\beta$ degrees of freedom, respectively. We stop the GPT-procedure when $-2(\mathcal{L}^k - \mathcal{L}^{k+1}) < \chi^2(\beta)$, where $\beta$ is replaced by the difference in the number of parameters (i.e., behaviors) between the two iterations.}
    , or when no negative cost column has been found at a given iteration. \rev{Further, we set the hyperparameters $\gamma=10$, and $\delta=20$ (as in \cite{jena2020partially}), which denote the number of behaviors to sample and the number of best ones to add to the master problem at each iteration, respectively.} For the estimaton of PCMC, we used the code provided by the authors\footnote{Code available at https://github.com/sragain/pcmc-nips.}.
    We refer the reader to Appendix \ref{ap:pcmc_implem} for more details on the implementation of the PCMC choice model.
    %
    
    
    \subsection{Numerical results on synthetic instances}
    \label{subsec:synthetic_exps}
    \rev{We start this section by describing our data generation procedure. We then proceed by characterizing the irrationality level of the generated instances, and by reporting the generalization performance of the various approaches on such instances.}
    
    \subsubsection{Data Generation.}
    \label{sec:datagen}
    We generate choice data samples according to two ground-truth models, specifically the Halo-MNL model proposed by \cite{HaloMNL} and the GSP model. Both of them allow us to control the amount of irrationality resulting in the generated instances and to investigate its impacts on the performance of the various approaches. For each ground truth model, instances were generated as follows:
    \begin{itemize}
        \item \textbf{Halo-MNL}: this choice model is parametrized by a pairwise interaction matrix $U$, whose diagonal terms $u_{ii}$ represent the item-specific utilities. Given the offer set $S$, the overall probability of choosing product $i$  is given by
            \begin{equation*}
                P(i|S) = \frac{\text{exp}({u_{ii} + \sum_{k \notin S} u_{ki}})}{\sum_{j\in S} \text{exp}({u_{jj} + \sum_{k \notin S} u_{kj}})} \text{ .}
            \end{equation*}
            It is easy to see that by setting to zero the off-diagonal terms of matrix $U$, we obtain an MNL model. Following \cite{DecisionForest-misic}, we draw the elements $u_{ii} \sim \text{Unif}[-1,1] $. We vary the irrationality of the instances by varying the number of pairwise interactions. Specifically, we generate instances where 0\%, 10\% and 25\% of the couples present a positive interaction, obtained by setting the corresponding off-diagonal terms to -1. We simulate both symmetric halo effects, where two products increase each other's attractiveness, and asymmetric halo effects, also known as \textit{decoy} effects, where only one of two products benefits from the presence of the other (the decoy) in the offer set. In order to investigate more complex interaction scenarios, we generalize the Halo-MNL model so as to include multiple customer segments, whose probability is drawn uniformly from the unit simplex. In our experiments, we have used either one or ten customer segments. We note again that when setting the number of pairwise interactions to zero, we end up obtaining rational instances generated under MNL and MMNL ground-truth models, depending on the number of customer segments, 1 and 10, respectively.
        \item \textbf{GSP}: We remind from Section \ref{sec:model} that a generalized stochastic preference is defined as $C(\sigma,i)$, where $\sigma$  is a ranking over the $N$ alternatives, and $i$ is the irrationality level of the customer type. Instances generated under this ground-truth model contain either 10 or 100 customer types, whose probabilities are randomly drawn from the unit simplex. For each instance, we consider 10\%, 20\% or 50\% of the customer types as irrational, meaning their index $i$ is greater than one.  Specifically, the irrationality level $i$ of each of these customer types was randomly chosen in $\{1,2,...,i_{max}\}$. We used $i_{max}=1,5,\text{ and }9$ to simulate various levels of irrationality. Rational instances have been obtained by setting the percentage of irrational behaviors to zero.
    \end{itemize}
    
    In all the experiments reported in this section, we used a number of products $N=10$, one of which represents the no-purchase option. 
    For each ground-truth model, we generate either $3,000$ or $50,000$ transactions, for a total of 10, 20 or 50 training offer sets. This simulates different amounts and diversity of training data. When using 50,000 transactions, in particular, the goal is to simulate the scenario in which we train the models based on empirical probabilities that are close to the true ones (i.e., those from the ground-truth model), and the effect of any sampling noise becomes negligible. This corresponds to the setting already used, for example, in \cite{bertsimas16-local_search} and \cite{DecisionForest-misic}, where choice models are trained on ground truth probabilities. We further assume that the number of transactions is equally distributed among the training offer sets, which all have dimension $|S| \geq 3$ and contain the no-purchase option. 
    
    \vspace{0.1cm}

    \subsubsection{Loss of Rationality.}
    \label{sec:lor}
       \begin{figure}[!htbp]
       \centering{
        \includegraphics[scale=0.6]{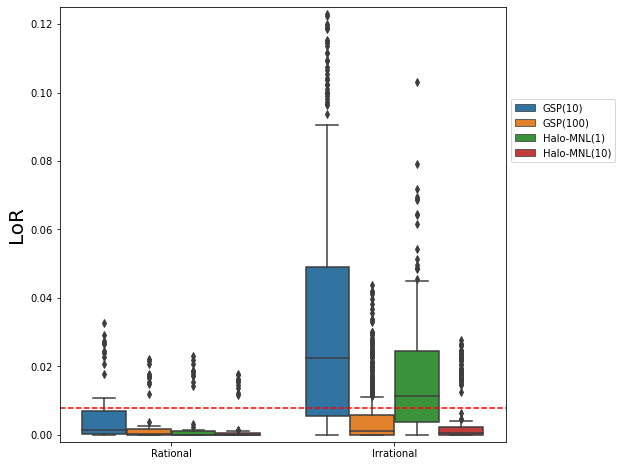}
        }  
        \caption{Distributions of \textit{Loss of Rationality} for generated instances, grouped by ground-truth models and number of customer behaviors.}
        \label{fig:Lor-distrib}
    \end{figure}

    We first investigate the level of \textit{irrationality} present in the generated instances. In line with the methodology proposed by \cite{LoR-jagabathula}, we fit the enumerative rank-based choice model, RB, to all our training instances. It is well known that any RUM choice model can be equivalently represented as a probability distribution over rankings of alternatives \citep{block1959random}. Thus, by fitting such model to a given instance, the resulting objective function indicates what the authors define as the 
    \textit{Loss of rationality} (LoR) of that instance, which can be interpreted as a measure of the minimum amount of choice data that cannot be explained by using any choice model belonging to the RUM family. Figure \ref{fig:Lor-distrib} reports the LoR value distributions over instances grouped by category (Rational and Irrational), ground-truth models (Halo-MNL and GSP) and number of customer types (in parenthesis). As already mentioned, rational instances for Halo-MNL(1) and Halo-MNL(10) correspond to instances generated under MNL and MMNL ground-truth models, respectively. Also, rational GSP ground-truth models are equivalent to Rank-Based models  with the same number of preference lists.
    %
    %
   
   It is interesting to note that the aggregation of a high number of irrational customer types seems to result into a rational choice behavior at the population level (see Halo-MNL(10) and GSP(100) in Figure \ref{fig:Lor-distrib}),  
   \rev{
   given that the individual, complex buying behavior becomes less apparent, and the collective (aggregated) transactions are easier approximated by simple choice rules.
   This is also connected to what \cite{comparativeBerbeglia} refers to as degree of \textit{consistency}. Instances with many customer types are said to be ``less consistent'', since the probability of being of a certain customer type is relatively low, and have been found to generalize better.
    }
   
   Figure \ref{fig:Lor-distrib} also reports a red dashed line, corresponding to a Loss of rationality of 0.008, which visually separates the generated instances based on their irrationality level. Essentially, rationally-generated instances tend to fall below this threshold, but also the irrationally-generated ones in which many customer types are aggregated. In the following, we interpret this value as a threshold to understand whether an instance contains significant amount of irrational choice behaviors. In Appendix \ref{ap:dataImpact_lor}, we further show that the LoR of a given instance can be impacted by other factors as well, such as the number of choice samples and offer sets available for training. 
   

    \vspace{0.1cm}
    
    \subsubsection{Generalization performances.}
       \label{subsubsec:genperf_synth}

    We now focus on the generalization performance of the various approaches when tested on new offer sets. This has been measured in terms of average $L_1$ error between the predicted probability distribution $\pmb{x}$ and the ground-truth probability distribution $\pmb{v}$ on new offer sets, and has been computed as
     \begin{equation}\label{eq:l1_expect_rand}
        L_1(\pmb{x}, \pmb{v}) = \frac{1}{|\mathcal{S}_{test}|} \sum_{S \in \mathcal{S}_{test}} \sum_{i \in S} \big|x_{i,S} - v_{i,S}\big| \text{,}
    \end{equation}
    where $\mathcal{S}_{test}$ is the collection of \textit{all} possible offer sets that have not been used for training\footnote{The total number of offer sets with dimension $3 \leq |S| \leq 10$ and containing the no-purchase option is equal to $502$. Hence, given $M=10, 20$ or $50$, $|\mathcal{S}_{test}| = 502- M$.}. 
    As noted in \cite{PCMC}, equation (\ref{eq:l1_expect}) can be interpreted as the expected $L_1$ prediction error given a randomly drawn offer set.

    Table \ref{tab:L1_err_unstruct} reports the $L_1$ test errors of each approach on sets of instances grouped by ground-truth models and number of customers types, indicated in parenthesis. 
    \begin{table}[!htpb]
        \centering
        \small
        \renewcommand{\arraystretch}{1.2}
        \begin{tabular}{lrrrrrrrr}																		
        \toprule																		
        		&	\% irrat	&	LoR	&	RB-R	&	GPT-R	&	GPT-I	&	GPT-IC	&	PCMC	&	Halo-MNL	\\
        \midrule
        	Irrational instances &		&		&		&	&		&  &		&		\\
        	\cmidrule{1-1}
        	Halo-MNL(1)	&	10	&	0.0103	&	0.2176	&	0.2305	&	0.1693	&	\textbf{0.1648}	&	0.2606	&	0.2028	\\
        	Halo-MNL(1)	&	25	&	0.0244	&	0.3229	&	0.3280	&	0.2843	&	0.2801	&	0.3804	&	\textbf{0.2370}	\\
        	        	&	(all) Mean 	&	0.0173	&	0.2702	&	0.2792	&	0.2268	&	0.2224	&	0.3205	&	\textbf{0.2199}	\\
        	        	\addlinespace
        	Halo-MNL(10)	&	10	&	0.0035	&	0.1462	&	0.1128	&	0.0995	&	\textbf{0.0979}	&	0.2196	&	0.1917	\\
        	Halo-MNL(10)	&	25	&	0.0044	&	0.1689	&	0.1574	&	0.1421	&	\textbf{0.1378}	&	0.2518	&	0.2038	\\
        		            &	(all) Mean 	&	0.0040	&	0.1575	&	0.1351	&	0.1208	&	\textbf{0.1179}	&	0.2357	&	0.1977	\\
        		           \addlinespace
        	GSP(10)	&	10	&	0.0171	&	0.2441	&	0.2208	&	\textbf{0.2171}	&	0.2337	&	0.4271	&	0.6258	\\
        	GSP(10)	&	20	&	0.0286	&	0.2805	&	0.2625	&	\textbf{0.2538}	&	0.2675	&	0.4624	&	0.6868	\\
        	GSP(10)	&	50	&	0.0573	&	0.3735	&	0.3651	&	\textbf{0.3388}	&	0.3563	&	0.5314	&	0.8100	\\
        	    	&	(all) Mean 	&	0.0343	&	0.2994	&	0.2828	&	\textbf{0.2699}	&	0.2858	&	0.4736	&	0.7075	\\
        	    	    \addlinespace
        	GSP(100)	&	10	&	0.0034	&	0.1679	&	\textbf{0.1409}	&	0.1440	&	0.1438	&	0.2890	&	0.2702	\\
        	GSP(100)	&	20	&	0.0043	&	0.1801	&	\textbf{0.1573}	&	0.1617	&	0.1587	&	0.3009	&	0.2737	\\
        	GSP(100)	&	50	&	0.0082	&	0.2160	&	0.2044	&	0.2046	&	\textbf{0.2030}	&	0.3394	&	0.3154	\\
        		&	(all) Mean 	&	0.0053	&	0.1880	&	\textbf{0.1675}	&	0.1701	&	0.1685	&	0.3098	&	0.2865	\\
        		        \addlinespace
        	(all) Mean 	& &	0.0170	&	0.2345	&	0.2196	&	\textbf{0.2058}	&	0.2096	&	0.3567	&	0.4083	\\
        	(all) Median& &	0.0046	&	0.2083	&	0.1882	&	0.1772	&	\textbf{0.1734}	&	0.3279	&	0.3078	\\
        	(all) Max   & &	0.2343	&	0.9454	&	0.7877	&	0.8087	&	\textbf{0.7835}	&	1.0035	&	1.5872	\\
        \midrule		
        Rational instances &		&		&		&	&		&  &		&		\\
        	\cmidrule{1-1}
        	MNL	&	-	&	0.0034	&	0.1227	&	\textbf{0.0916}	&	0.0933	&	0.0918	&	0.1404	&	0.1772	\\
        	MMNL	&	-	&	0.0027	&	0.1294	&	\textbf{0.0724}	&	0.0762	&	0.0760	&	0.1529	&	0.1823	\\
        	RB(10)	&	-	&	0.0060	&	0.1622	&	\textbf{0.1468}	&	0.1636	&	0.1812	&	0.3841	&	0.5539	\\
        	RB(100)	&	-	&	0.0033	&	0.1559	&	\textbf{0.1251}	&	0.1317	&	0.1282	&	0.2793	&	0.2498	\\
                        \addlinespace
        	(all) Mean	&		&	0.0039	&	0.1425	&	\textbf{0.1090}	&	0.1162	&	0.1193	&	0.2392	&	0.2908	\\
        	(all) Median	&		&	0.0003	&	0.1361	&	\textbf{0.1012}	&	0.1097	&	0.1088	&	0.2310	&	0.2291	\\
        	(all) Max	&		&	0.0327	&	0.4697	&	0.4117	&	0.4211	&	\textbf{0.4075}	&	0.6795	&	1.2308	\\
        \bottomrule																		
        \end{tabular}																		
        \caption{$L_1$ test errors for each approach on Irrational instances grouped by ground-truth model, number of customer types(in parenthesis) and percentage of irrational behaviors/interactions used to generate the data.}
        \label{tab:L1_err_unstruct}
    \end{table}

    \rev{We first focus on the set of irrational instances. It is possible to observe how the performance of the rational choice models, RB-R and GPT-R, deteriorates as the Loss of Rationality increases. Specifically, this is the case for Halo-MNL(1) and GSP(10) instances where, among the rank-based approaches, GPT-IC and GPT-I offer the best predictive accuracy, respectively. GPT-IC favourably compares also to the Halo-MNL for small enough percentage of positive, pairwise interactions among products. On average, however, the Halo-MNL choice model manages to well approximate the data-generating process on Halo-MNL(1) instances, therefore outperforming all other approaches. We notice that GPT-IC tends to outperform GPT-I on all classes of instances, with the exception of those in  GSP(10). On these instances, GPT-IC does actually not improve over GPT-R, unless there is a high percentage of irrational behaviors. Such results are in line with the behavioral assumption intrinsic to GPT-IC, which we proposed to explicitly prioritize sparse, low-order interactions among products, and may thus fail to generalize well on those scenarios where customers consistently (i.e., with high probability) show high orders of substitution and halo effects among products. However, we deem such interactions less likely to happen in practice. 
    Our results on real-world data (see Section \ref{subsec:real_exps}) seem to confirm this claim. The good median and maximum test errors of GPT-IC further confirm the robustness of the underlying selection criterion to identify customer behaviors that tend to generalize well.}
    
    \rev{  We now discuss the results for rational instances, where the flexibility of irrational approaches may increase the risk of overfitting (see also Appendix \ref{ap:learning_stats}) when compared to GPT-R. 
    \sout{This also} All GPT-based algorithms  favorably compare with RB-R, confirming the results of \cite{jena2020partially} on the generalization power of partially-ranked lists and that enumerating all possible fully-ranked preferences for training the RB model increases the risk of overfitting the training set \citep{Vulcano-marketDiscovery}. 
    This supports the hypothesis that adding only relevant types to the estimated choice model is crucial for its generalization performance. 
    Also note that the same relative performance between GPT-I and GPT-IC observed on irrational instances can be noticed in the rational case as well. Specifically, GPT-IC outperforms GPT-I, on average, on all instances with the exception of RB(10) ones, i.e., those where customers consistently exhibit high order of substitutions among products.
    Finally, we emphasize that both irrational GPT variants significantly outperform the PCMC and Halo-MNL models on such rational instances.
    }

\begin{figure}[!htpb]
         \centering
         \includegraphics[scale=0.5]{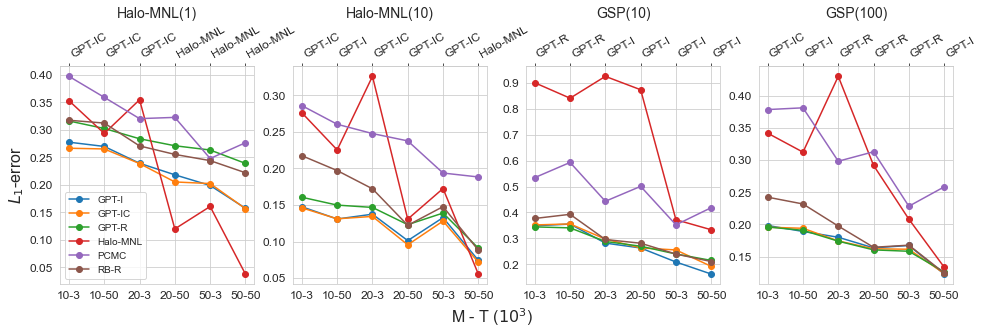}
         \caption{Average L1 test error of the various approaches on Irrational instances. The bottom axis reports the number of unique assortments $M$ and transactions $T$ (in $10^3$ units) available for training. The top axis shows, for each data-regime setting, the best performing approach.}
         \label{fig:dataav_irrat}
     \end{figure}

\begin{figure}[!htpb]
         \centering
         \includegraphics[scale=0.5]{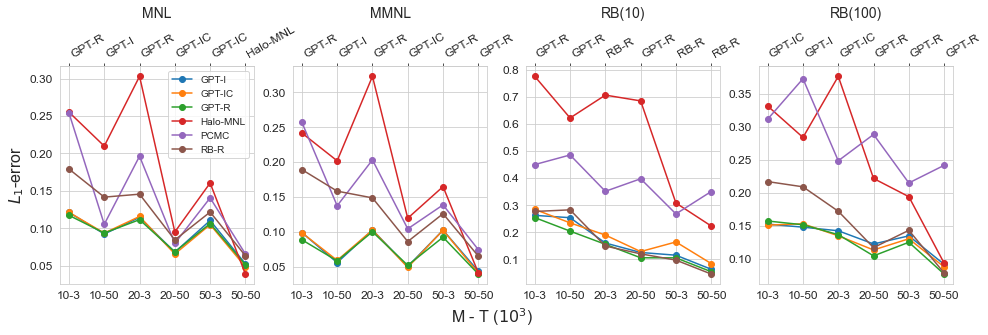}
         \caption{Average L1 test error of the various approaches on rational instances. The bottom axis reports the number of unique assortments $M$ and transactions $T$ (in $10^3$ units) available for training. The top axis shows, for each data-regime setting, the best performing approach.}
         \label{fig:dataav_rat}
     \end{figure}

    \paragraph{Impact of the amount of available data}
    \rev{We now test the robustness of the various approaches under different amounts of training data. In particular, Figure \ref{fig:dataav_irrat} and Figure \ref{fig:dataav_rat} report, for each approach, the corresponding average test error over irrational and rational instances, respectively, under different data-regime settings ``$T$-$M$'', where $T$ represents the number of transactions and $M$ the number of assortments available for training. Each plot also reports, on the top axis, the best performing model under each data-regime scenario.}
    
    \rev{Both figures exhibit common trends in the the relative performance of the approaches. In particular, the Halo-MNL model tends to require significant amounts of data in order to become competitive with (or outperform) the rank-based approaches. These, in contrast, show relative stable performance and steadily improve with more available training data. They also consistently outperform the PCMC model.}
    
    \rev{Among the rank-based approaches, in line with our discussion of Table \ref{tab:L1_err_unstruct}, we observe that GPT-IC tends to perform the best on Halo-MNL instances. Also, while being competitive on GSP(100) and RB(100) instances, it is generally outperformed by other rank-based approaches on GSP(10) and RB(10) instances, which are characterized by higher orders of interactions among products.}

    \vspace{0.1cm}
    \rev{We conclude this section by referring the interested reader to Appendix \ref{ap:add_res} for additional numerical results. In particular, Appendix \ref{ap:learning_stats} provides statistics describing the choice models learned by GPT-based approaches on classes of instances characterized by different levels of irrationality. The results therein also highlight the computational effectiveness of GPT-based approaches, which can be estimated in less than two seconds, on average, on our synthetic instances.
    In Appendix \ref{ap:gsp_inst}, we provide a more refined view of the impact of the irrationality level of GSP customer types on the predictive accuracy of the various approaches.  Finally, Appendix \ref{ap:halo_SymmVsAsymm} investigates the generalization performance of the implemented algorithms on Halo-MNL instances separated by type of positive interaction, i.e., symmetric or asymmetric, in the ground-truth model.}

\subsection{Numerical results on the IRI Academic dataset}
\label{subsec:real_exps}
In this section, we test all approaches on the IRI Academic data-set \citep{bodea2009data}, a real-world data-set from the retail sector. The data consists of transactions from grocery and drug store chains located in 47 markets in the USA.  We consider transactions corresponding to 29 product categories in total, which we report in Table \ref{tab:irires}. 

\subsubsection{Dataset pre-processing. }
\rev{Each transaction record in the dataset contains informations regarding the week and store in which the transaction occured, and the purchased item, uniquely identified by its Universal Product Code (UPC).}
\rev{In order to tackle both the sparsity and large volume of the data, we follow the same pre-processing steps outlined in \cite{LoR-jagabathula}. Specifically, we start by considering the subset of transactions  corresponding to the first two weeks of year 2007. We then aggregate products by their UPC-vendor code. In other words, transactions where items with the same UPC-vendor code were bought, are attributed to the same \textit{vendor-item}. This kind of aggregation is common in the marketing literature in order to reduce the sparsity of the data \cite[see, e.g.,][]{bronnenberg2004market}. We then proceed by considering the nine most popular vendor-items, and further aggregate the remaining ones into a \textit{no-purchase} option, therefore obtaining ten alternatives in total.}

\rev{The assortment of products shown to the customer at the moment a transaction occurred is not given in the data. Hence, for each transaction $t=1,\dots,T$ occurring in week $w_t$ at the store $s_t$, where product $j_t$ was bought,  we assume the assortment shown to the customer consists of the set products sold at least once in that store-week combination, i.e., $S_t = \bigcup_{t'=1}^T\{j_{t'} : w_{t'}=w_t, z_{t'}=z_t\}$. At the end of this step we obtain a final list of transactions  $\mathcal{T}=\{(S_t, j_t)\}_{t=1}^T$.
}

\rev{
\subsubsection{Generalization performance. }
Following the experimental setup of \cite{DecisionForest-misic}, we separately test the various approach on each of the 29 product categories using 5-folds cross validation. In particular, after considering a product category with transactions data spanning  a collection $\mathcal{S}=\{S_1, \dots S_M\}$ of $M$ unique assortments, we partition $\mathcal{S}$ into five (approximately) equally sized collections of assortments $\mathcal{S}_1, \dots \mathcal{S}_5$. For each set $\mathcal{S}_i, i=1,\dots 5$,  let $\mathcal{T}_i$ denote the set of transactions where an assortment belonging to $\mathcal{S}_i$ was shown to the customers, i.e.,$ \mathcal{T}_i=\{(S_t, j_t) : S_t \in \mathcal{S}_i\}$ . We then run 5 separate experiments where transactions $\mathcal{T}_{train} = \mathcal{T}\setminus \mathcal{T}_i$ are used for training while transactions  $\mathcal{T}_i$ are used for testing the predictive accuracy of the algorithms on new assortments.
}

\rev{We measure the predictive accuracy in terms of expected $L_1$ error, over unseen assortments. We modify equation (\ref{eq:l1_expect_rand}) to account for the fact that each test assortment is now associated to a certain number of transactions $T_S=\sum_{t=1}^T \mathbbm{1}[S_t=S]$. We thus compute the test error as follows:}
    \begin{equation}\label{eq:l1_expect}
        L_1(\pmb{x}, \pmb{v}) = \frac{1}{|\mathcal{T}_{test}|} \sum_{S \in \mathcal{S}_{test}} T_S \sum_{i \in S} \big|x_{i,S} - v_{i,S}\big| \text{.}
    \end{equation}

\begin{table}[!htpb]
    \centering
    \small
    \begin{tabular}{lrrrrrrrrr}																	
\toprule																	
Product Category	&	$M$	&	$T$	&	RB-R	&	GPT-R	&	GPT-I	&	GPT-IC	&	PCMC	&	Halo-MNL	\\
\midrule																	
Beer	&	55	&	380,932	&	0.2727	&	0.2729	&	\textbf{0.1272}	&	0.1619	&	0.2136	&	0.1498	\\
Blades	&	57	&	92,404	&	0.0656	&	0.0565	&	0.0591	&	0.0581	&	0.0732	&	\textbf{0.0429}	\\
Carbonated Beverages	&	31	&	721,506	&	0.1326	&	0.1284	&	0.1365	&	\textbf{0.1043}	&	0.1403	&	0.1249	\\
Cigarets	&	68	&	249,668	&	0.1396	&	0.1428	&	\textbf{0.0677}	&	0.0726	&	0.0971	&	0.0765	\\
Coffee	&	47	&	372,536	&	0.1636	&	0.1681	&	0.1944	&	\textbf{0.1581}	&	0.1808	&	0.1773	\\
Cold Cereal	&	15	&	577,236	&	0.1351	&	0.1328	&	0.1021	&	0.0817	&	0.0739	&	\textbf{0.0663}	\\
Deodorant	&	45	&	271,286	&	0.0781	&	0.0791	&	0.0713	&	\textbf{0.0656}	&	0.0856	&	0.0847	\\
Diapers	&	18	&	143,055	&	0.1204	&	0.1277	&	0.1305	&	\textbf{0.1120}	&	0.3627	&	0.1987	\\
Facial Tissue	&	43	&	73,806	&	0.1185	&	0.1078	&	\textbf{0.0935}	&	0.0942	&	0.1832	&	0.1173	\\
Frozen Dinners	&	30	&	979,936	&	0.1952	&	\textbf{0.1845}	&	0.2352	&	0.2081	&	0.2213	&	0.2720	\\
Frozen Pizza	&	61	&	292,878	&	0.1406	&	0.1427	&	0.1070	&	0.1038	&	0.1184	&	\textbf{0.1027}	\\
Household Cleaners	&	19	&	282,981	&	0.1467	&	0.1410	&	\textbf{0.1299}	&	0.1361	&	0.1313	&	0.1527	\\
Hotdogs	&	100	&	101,624	&	0.1961	&	0.1920	&	0.2064	&	\textbf{0.1733}	&	0.1983	&	0.1884	\\
Laundry Detergent	&	56	&	238,163	&	0.1425	&	\textbf{0.1363}	&	0.1491	&	0.1479	&	0.1490	&	0.1453	\\
Margarine/Butter	&	18	&	140,969	&	0.1750	&	0.1583	&	0.1120	&	0.1264	&	\textbf{0.0975}	&	0.1371	\\
Mayonnaise	&	48	&	97,282	&	0.0776	&	\textbf{0.0756}	&	0.0858	&	0.0818	&	0.0914	&	0.0914	\\
Milk	&	49	&	240,691	&	0.1655	&	0.1618	&	0.1439	&	0.1493	&	0.1359	&	\textbf{0.1243}	\\
Mustard/Ketchup	&	44	&	134,800	&	0.1204	&	0.1184	&	\textbf{0.0973}	&	\textbf{0.0973}	&	0.1005	&	0.1073	\\
Paper Towels	&	40	&	82,636	&	0.1335	&	\textbf{0.1278}	&	0.1335	&	\textbf{0.1278}	&	0.1439	&	0.1654	\\
Peanut Butter	&	51	&	108,770	&	0.1016	&	0.0991	&	0.1021	&	\textbf{0.0979}	&	0.1347	&	0.1058	\\
Salty Snacks	&	39	&	736,148	&	\textbf{0.1300}	&	0.1381	&	0.1742	&	0.1420	&	0.1378	&	0.1362	\\
Shampoo	&	66	&	290,429	&	0.1215	&	0.1184	&	0.1096	&	\textbf{0.0995}	&	0.1198	&	0.1082	\\
Soup	&	24	&	905,541	&	0.1053	&	\textbf{0.1037}	&	0.1363	&	0.1199	&	0.1266	&	0.1412	\\
Spaghetti/Italian Sauce	&	38	&	276,860	&	0.2017	&	0.2074	&	0.2556	&	0.2049	&	\textbf{0.1857}	&	0.2100	\\
Sugar Substitutes	&	64	&	53,834	&	0.0800	&	0.0753	&	0.0724	&	0.0723	&	0.0732	&	\textbf{0.0626}	\\
Toilet Tissue	&	27	&	112,788	&	0.1460	&	0.1400	&	0.1284	&	\textbf{0.1113}	&	0.1477	&	0.1518	\\
Toothbrush	&	114	&	197,676	&	0.1425	&	0.1411	&	0.1145	&	\textbf{0.1128}	&	0.1313	&	0.1143	\\
Toothpaste	&	42	&	238,271	&	0.0693	&	0.0698	&	\textbf{0.0644}	&	0.0650	&	0.0742	&	0.0734	\\
Yogurt	&	43	&	499,203	&	0.1584	&	0.1703	&	0.1605	&	0.1468	&	\textbf{0.1278}	&	0.1712	\\
\midrule
Mean 	&	-	&	-	&	0.1371	&	0.1351	&	0.1276	&	\textbf{0.1184}	&	0.1399	&	0.1310	\\
GPT-IC Improvement 	&	-	&	-	&	13.7\%	&	12.4\%	&	7.2\%	&	-	&	15.4\%	&	9.7\% \\	
Median	&	-	&	-	&	0.1324	&	0.1283	&	0.1200	&	\textbf{0.1086}	&	0.1271	&	0.1207	\\
Max	&	-	&	-	&	0.3717	&	0.3513	&	0.3438	&	\textbf{0.3110}	&	0.7175	&	0.3867	\\
Nb Best	&	-	&	-	&	1	&	5	&	6	&	\textbf{11}	&	3	&	5	\\
\bottomrule																	
\end{tabular}																											
    \caption{Comparison of the various approaches on the IRI Academic Dataset. The reported metric is the average, $L_1$ error, obtained using 5-folds cross-validation for for each product category.}
    \label{tab:irires}
\end{table}

\rev{Table \ref{tab:irires} reports the average test errors of the various approaches on each product category. Columns $M$ and $T$ indicate the total number of unique assortments and transactions, respectively, available for each product category. We further report for each approach the mean, median and maximum test error over $all$ experiments. The row ``GPT-IC Improvement'' reports the percentage improvement of GPT-IC over each approach, while ``Nb Best'' denotes the number of product categories in which the corresponding approach achieves the best average predictive accuracy.}

\rev{We observe that GPT-IC significantly outperforms all other approaches according to all metrics on this dataset. Such results confirm that prioritizing customer behaviors with a relatively small number of relevant, strictly ranked products, may enhance predictive accuracy in practice. Specifically, GPT-IC achieves a 7.2\% improvement in predictive accuracy with respect to GPT-I, on average, and a 12.4\% improvement with respect to GPT-R, the best RUM baseline.}
\rev{In this context, one may wonder whether the predictive accuracy of GPT-R can be improved by adopting the new customer selection rule of GPT-IC (see Section \ref{sec:newrule}). We therefore tested such a configuration on the IRI data-set. However, our experiments did not show any improved test errors, confirming that the new dominance rule is indeed helpful only when aiming at estimating irrational customer preference sequences.}

\begin{figure}[!htpb]
        \centering
        \includegraphics[scale=0.55]{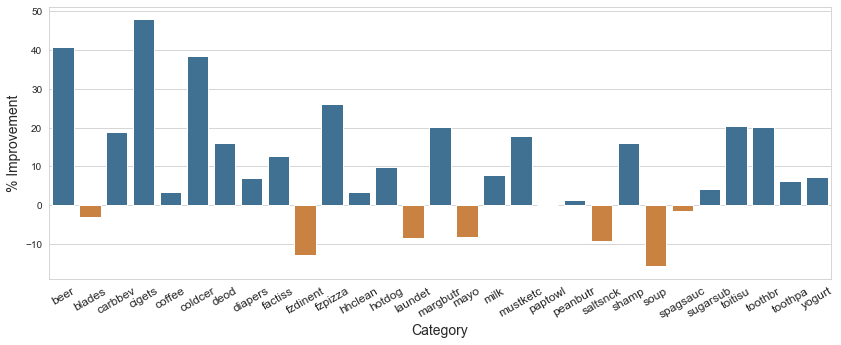}
        \caption{Percentage Improvement of GPT-IC over the \textit{best} RUM baseline (between RB-R and GPT-R) in terms of $L_1$ test error on each Product Category.}
        \label{fig:gpticVSrum}
    \end{figure}

    \rev{Figure \ref{fig:gpticVSrum} provides a clearer picture of the improvement one may achieve by going beyond RUM on each product category. In particular, for each category we compare GPT-IC with the \textit{best} performing RUM baseline, namely RB-R and GPT-R, and then report the average percentage improvement in predictive accuracy. Notably, such improvements can be as high as 48\% for the category of cigarettes, and close to 40\% for the beer and coffee product categories. We also note that for those product categories where the decrease in accuracy is the most significant, such as frozen dinner, laundry and detergent, mayonnaise, salty snacks and soup, none of the irrational baseline we tested was able to outperform the RUM ones. This shows that either such products categories do not contain significant levels of irrationality, or not enough data is available to learn such complex product interactions.}
  
\begin{figure}[!htpb]
        \centering
        \includegraphics[scale=0.45]{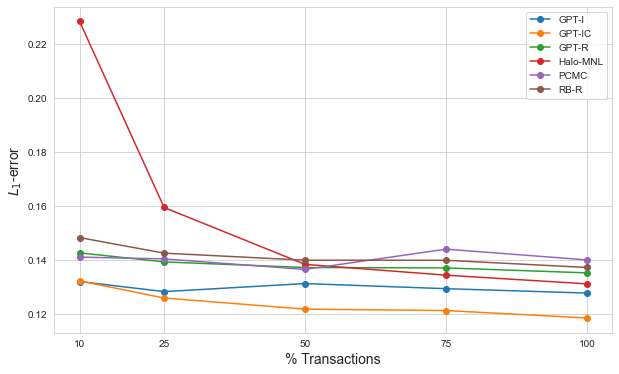}
        \caption{$L_1$ test error of the various approaches, averaged over all product categories, for various amounts of training transactions data.}
        \label{fig:Learningcurves}
    \end{figure}

    \vspace{0.1cm}
    \paragraph{Impact of the amount of available data. }
    \rev{
    In this section we investigate how the performance of the various approaches is affected by the amount of data available for training. In particular, Figure \ref{fig:Learningcurves} reports  the average test error over all products categories when different amount of transactions are used for training. Specifically, for each of product category, we perform 5-fold cross validations as described above and, for each experiment $i=1,\dots,5$, we select 10\%, 25\%, 50\%, 75\% and 100\% of the transactions in $\mathcal{T}_{train}$, respectively.}
    
    \rev{In line with our results on synthetic instances (see Section \ref{subsec:synthetic_exps}), Halo-MNL needs significant amounts of data to be competitive with rank-based approaches, only outperforming GPT-R when 75\% or more of the available transactions are used for training. In contrast, the GPT-based approaches seem to be quite data-efficient, with relatively stable performances even for the case where only 10\% of the transactions are used for training. Interestingly, the performance of GPT-I slightly deteriorates when moving from 25\% to 50\% of the transaction being used. This may be due to the fact that, for several product categories, most of the transactions concern a relatively small number of assortments. Hence, cases with only 10\% of training transactions tend to exclude ``noisy'' assortments for which only few transactions (as few as one or two) are available, and which may cause GPT-I to infer spurious product interactions. GPT-IC, on the other hand, does not seem to suffer from such issue, since  it prioritizes low-order interactions, which require less data to be learned effectively.}

\section{Conclusion}
\label{sec:conclusion}

\rev{
In this paper, we have proposed a discrete choice-model that is sufficiently flexible to capture rational and irrational choice-behavior, together with a computationally- and data-efficient estimation method. The resulting models have been shown to overfit less and generalize well when compared to existing benchmarks.

In particular, we extend the \textit{Generalized Stochastic Preference} choice model introduced by \cite{gsp-berbeglia2018} by adapting partially-ranked preference sequences \citep{jena2020partially}, which enables us to estimate the model via column generation.
In an effort to prevent additional overfitting induced by the more general irrational preference sequences, and linked to the concept of consideration sets from the marketing literature, we  propose a new criterion to select relevant customer types, which are more likely to generalize well in practice.

Our experiments on a popular and extensive set of real-world data has shown that our proposed models have a variety of advantages. 
First, the use of the new selection rule further improves the generalization accuracy of our irrational choice-model on unseen offer sets by 7.2\%, on average. In contrast, using this selection rule in the context of the purely rational choice-model did not result in improvements, confirming that this selection rule is indeed a contribution particular to the irrational case.
}
\rev{
Second, with respect to the rational baseline models, our irrational model boosts predictive accuracy by 12.4\%, on average, and for some categories up to 48\%.
On (synthetic) rational RUM instances, our irrational choice-models provide stable results, only slightly inferior to the rational baseline models, while the irrational baseline models showed significantly worse performance.
Third, our models have shown to be data-efficient, providing a higher predictive accuracy then the benchmarks when little training data is available.

Finally, an appealing feature of our approach is that, within the same framework, it is possible to exploit the explanatory power of (i) partially-ranked preference lists \citep[which can theoretically represent any RUM choice model, see, e.g.,][]{farias13} and (ii) irrational behaviors, which can significantly enhance accuracy on irrational instances. Using the new selection rule, our approach is thus capable of providing accurate estimates of product demands on both rational and irrational instances, circumventing the need for effective model selection criteria.
}

\ACKNOWLEDGMENT{We would like to acknowledge the  Wharton Customer Analytics (WCA) research group and the  Majid Al Futtaim (MAF) data \& analytics team for the insightful discussions.}

\renewcommand\bibname{REFERENCES}
\bibliographystyle{ormsv080}  
\bibliography{clean_biblio}

\begin{thebibliography}{49}
\expandafter\ifx\csname natexlab\endcsname\relax\def\natexlab#1{#1}\fi
\expandafter\ifx\csname url\endcsname\relax
  \def\url#1{{\tt #1}}\fi
\expandafter\ifx\csname urlprefix\endcsname\relax\def\urlprefix{URL }\fi
\expandafter\ifx\csname urlstyle\endcsname\relax
  \expandafter\ifx\csname doi\endcsname\relax
  \def\doi#1{doi:\discretionary{}{}{}#1}\fi \else
  \expandafter\ifx\csname doi\endcsname\relax
  \def\doi{doi:\discretionary{}{}{}\begingroup \urlstyle{rm}\Url}\fi \fi

\bibitem[{Aouad et~al.(2021)Aouad, Farias, and Levi}]{aouad2015assortment}
Aouad, Ali, Vivek Farias, Retsef Levi. 2021.
\newblock Assortment optimization under consider-then-choose choice models.
\newblock {\it Management Science\/} {\bf 67}(6) 3368--3386.

\bibitem[{Ariely(2008)}]{ariely2008predictably}
Ariely, D. 2008.
\newblock {\it Predictably irrational\/}.
\newblock HarperCollins, New York, NY, US.

\bibitem[{Berbeglia(2018)}]{gsp-berbeglia2018}
Berbeglia, G. 2018.
\newblock {The generalized stochastic preference choice model}.
\newblock {\it arXiv preprint arXiv:1803.04244\/} .

\bibitem[{Berbeglia et~al.(2018)Berbeglia, Garassino, and
  Vulcano}]{comparativeBerbeglia}
Berbeglia, G., A.~Garassino, G.~Vulcano. 2018.
\newblock A comparative empirical study of discrete choice models in retail
  operations.
\newblock {\it Available at SSRN 3136816\/} .

\bibitem[{Bertsimas and Mi\v{s}ic(2016)}]{bertsimas16-local_search}
Bertsimas, D., V.~Mi\v{s}ic. 2016.
\newblock Data-driven assortment optimization.
\newblock {\it Tech. report, Massachusetts Institute of Technology\/} .

\bibitem[{Blanchet et~al.(2016)Blanchet, Gallego, and Goyal}]{Blanchet2016}
Blanchet, J., G.~Gallego, V.~Goyal. 2016.
\newblock {A Markov Chain Approximation to Choice Modeling}.
\newblock {\it Operations Research\/} {\bf 64}(4) 886--905.

\bibitem[{Block and Marschak(1959)}]{block1959random}
Block, H.D., J.~Marschak. 1959.
\newblock Random orderings and stochastic theories of response.
\newblock Tech. Rep.~66, Cowles Foundation for Research in Economics, Yale
  University.

\bibitem[{Bodea et~al.(2009)Bodea, Ferguson, and Garrow}]{bodea2009data}
Bodea, Tudor, Mark Ferguson, Laurie Garrow. 2009.
\newblock Data set—choice-based revenue management: Data from a major hotel
  chain.
\newblock {\it Manufacturing \& Service Operations Management\/} {\bf 11}(2)
  356--361.

\bibitem[{Bronnenberg and Mela(2004)}]{bronnenberg2004market}
Bronnenberg, Bart~J, Carl~F Mela. 2004.
\newblock Market roll-out and retailer adoption for new brands.
\newblock {\it Marketing Science\/} {\bf 23}(4) 500--518.

\bibitem[{Busemeyer and Townsend(1993)}]{Busemeyer1993}
Busemeyer, J~R., J~T. Townsend. 1993.
\newblock {Decision field theory: A dynamic-cognitive approach to decision
  making in an uncertain environment}.
\newblock {\it Psychological Review\/} {\bf 100}(3) 432--459.

\bibitem[{Chen et~al.(2019)Chen, Gallego, and Tang}]{DecisionForest-gallego}
Chen, N., G.~Gallego, Z.~Tang. 2019.
\newblock The use of binary choice forests to model and estimate discrete
  choice models.
\newblock {\it Available at SSRN 3430886\/} .

\bibitem[{Chen and Mi\v{s}ic(2019)}]{DecisionForest-misic}
Chen, Y., V.~Mi\v{s}ic. 2019.
\newblock Decision forest: A nonparametric approach to modeling irrational
  choice.
\newblock {\it arXiv preprint arXiv:1904.11532\/} .

\bibitem[{Echenique and Saito(2019)}]{echenique2015general}
Echenique, F., K.~Saito. 2019.
\newblock General luce model.
\newblock {\it Economic Theory\/} {\bf 68}(4) 811--826.

\bibitem[{Echenique et~al.(2018)Echenique, Saito, and
  Tserenjigmid}]{percp-adjusted-luce}
Echenique, F., K.~Saito, G.~Tserenjigmid. 2018.
\newblock The perception-adjusted luce model.
\newblock {\it Mathematical Social Sciences\/} {\bf 93} 67--76.

\bibitem[{Farias et~al.(2013)Farias, Jagabathula, and Shah}]{farias13}
Farias, V.~F., S.~Jagabathula, D.~Shah. 2013.
\newblock A nonparametric approach to modeling choice with limited data.
\newblock {\it Management Science\/} {\bf 59}(2) 305--322.

\bibitem[{Feng et~al.(2018)Feng, Li, and Wang}]{Feng2017}
Feng, G., X.~Li, Z.~Wang. 2018.
\newblock On substitutability and complementarity in discrete choice models.
\newblock {\it Operations Research Letters\/} {\bf 46}(1) 141 -- 146.

\bibitem[{Gallego et~al.(2014)Gallego, Ratliff, and Shebalov}]{GAM}
Gallego, G., R.~Ratliff, S.~Shebalov. 2014.
\newblock {A general attraction model and sales-based linear program for
  network revenue management under customer choice}.
\newblock {\it Operations Research\/} {\bf 63}(1) 212--232.

\bibitem[{Hauser et~al.(2010)Hauser, Toubia, Evgeniou, Befurt, and
  Dzyabura}]{hauser2010considreview}
Hauser, John~R, Olivier Toubia, Theodoros Evgeniou, Rene Befurt, Daria
  Dzyabura. 2010.
\newblock Disjunctions of conjunctions, cognitive simplicity, and consideration
  sets.
\newblock {\it Journal of Marketing Research\/} {\bf 47}(3) 485--496.

\bibitem[{Hauser and Wernerfelt(1990)}]{hauser1990evaluation}
Hauser, John~R, Birger Wernerfelt. 1990.
\newblock An evaluation cost model of consideration sets.
\newblock {\it Journal of consumer research\/} {\bf 16}(4) 393--408.

\bibitem[{Honhon et~al.(2012)Honhon, Jonnalagedda, and Pan}]{Honhon2012}
Honhon, D., S.~Jonnalagedda, X.~A. Pan. 2012.
\newblock {Optimal Algorithms for Assortment Selection Under Ranking-Based
  Consumer Choice Models}.
\newblock {\it Manufacturing {\&} Service Operations Management\/} {\bf 14}(2)
  279--289.

\bibitem[{Huber et~al.(1982)Huber, Payne, and
  Puto}]{attreffect2_huber1982adding}
Huber, J., J.~W. Payne, C.~Puto. 1982.
\newblock Adding asymmetrically dominated alternatives: Violations of
  regularity and the similarity hypothesis.
\newblock {\it Journal of consumer research\/} {\bf 9}(1) 90--98.

\bibitem[{Huber and Puto(1983)}]{decoy_puto}
Huber, Joel, Christopher Puto. 1983.
\newblock Market boundaries and product choice: Illustrating attraction and
  substitution effects.
\newblock {\it Journal of consumer research\/} {\bf 10}(1) 31--44.

\bibitem[{Iyengar and Lepper(2000)}]{Iyengar2000}
Iyengar, S.~S., M.~R. Lepper. 2000.
\newblock {When choice is demotivating: can one desire too much of a good
  thing?}
\newblock {\it Journal of personality and social psychology\/} {\bf 79}(6)
  995--1006.

\bibitem[{Jagabathula and Rusmevichientong(2019)}]{LoR-jagabathula}
Jagabathula, S., P.~Rusmevichientong. 2019.
\newblock The limit of rationality in choice modeling: Formulation,
  computation, and implications.
\newblock {\it Management Science\/} {\bf 65}(5) 2196--2215.

\bibitem[{Jagabathula et~al.(2019)Jagabathula, Mitrofanov, and
  Vulcano}]{jagabathula2019inferring}
Jagabathula, Srikanth, Dmitry Mitrofanov, Gustavo Vulcano. 2019.
\newblock Inferring consideration sets from sales transaction data.
\newblock {\it NYU Stern School of Business\/} .

\bibitem[{Jena et~al.(2020)Jena, Lodi, Palmer, and Sole}]{jena2020partially}
Jena, Sanjay~Dominik, Andrea Lodi, Hugo Palmer, Claudio Sole. 2020.
\newblock A partially ranked choice model for large-scale data-driven
  assortment optimization.
\newblock {\it Informs Journal on Optimization\/} {\bf 2}(4) 297--319.

\bibitem[{Kleinberg et~al.(2017)Kleinberg, Mullainathan, and
  Ugander}]{KleinbergMU17}
Kleinberg, J., S.~Mullainathan, J.~Ugander. 2017.
\newblock Comparison-based choices.
\newblock {\it Proceedings of the 2017 ACM Conference on Economics and
  Computation\/}. 127--144.

\bibitem[{Luce(1959)}]{Luce1959}
Luce, R.~D. 1959.
\newblock {\it Individual choice behavior\/}.
\newblock John Wiley \& Sons, New York.

\bibitem[{Maragheh et~al.(2018)Maragheh, Chronopoulou, and Davis}]{HaloMNL}
Maragheh, R.~Y., A.~Chronopoulou, J.~M. Davis. 2018.
\newblock A customer choice model with halo effect.
\newblock {\it arXiv preprint arXiv:1805.01603\/} .

\bibitem[{Mottini and Acuna-Agost(2017)}]{PointerNet}
Mottini, A., R.~Acuna-Agost. 2017.
\newblock Deep choice model using pointer networks for airline itinerary
  prediction.
\newblock {\it Proceedings of the 23rd ACM SIGKDD International Conference on
  Knowledge Discovery and Data Mining\/}. 1575--1583.

\bibitem[{Nocedal and Wright(2006)}]{nocedal2006numerical}
Nocedal, J., S.~Wright. 2006.
\newblock {\it Numerical optimization\/}.
\newblock Springer Series in Operations Research, Springer, New York, NY, US.

\bibitem[{Osogami and Otsuka(2014)}]{osogami-rbm}
Osogami, T., M.~Otsuka. 2014.
\newblock Restricted boltzmann machines modeling human choice.
\newblock {\it Advances in Neural Information Processing Systems\/}. 73--81.

\bibitem[{Pfannschmidt et~al.(2019)Pfannschmidt, Gupta, and
  H{"u}llermeier}]{FATE}
Pfannschmidt, K., P.~Gupta, E.~H{"u}llermeier. 2019.
\newblock Learning choice functions.
\newblock {\it arXiv preprint arXiv:1901.10860\/} .

\bibitem[{Ragain and Ugander(2016)}]{PCMC}
Ragain, S., J.~Ugander. 2016.
\newblock Pairwise choice markov chains.
\newblock {\it Advances in Neural Information Processing Systems\/}.
  3198--3206.

\bibitem[{Rieskamp et~al.(2006)Rieskamp, Busemeyer, and Mellers}]{Rieskamp2006}
Rieskamp, J., J.~R. Busemeyer, B.~A. Mellers. 2006.
\newblock {Extending the Bounds of Rationality: Evidence and Theories of
  Preferential Choice}.
\newblock {\it Journal of Economic Literature\/} {\bf 44}(3) 631--661.

\bibitem[{Roe et~al.(2001)Roe, Busemeyer, and Townsend}]{Roe2001}
Roe, R.~M., J.~R. Busemeyer, J.~T. Townsend. 2001.
\newblock {Multialternative decision field theory: a dynamic connectionist
  model of decision making.}
\newblock {\it Psychological review\/} {\bf 108}(2) 370--392.

\bibitem[{Rooderkerk et~al.(2011)Rooderkerk, {Van Heerde}, and
  Bijmolt}]{Rooderkerk2011}
Rooderkerk, R.~P., H.~J. {Van Heerde}, T.~H.A Bijmolt. 2011.
\newblock {Incorporating Context Effects Into a Choice Model}.
\newblock {\it Journal of Marketing Research\/} {\bf 48}(4) 767--780.

\bibitem[{Rosenfeld et~al.(2020)Rosenfeld, Oshiba, and
  Singer}]{rosenfeld2019predicting}
Rosenfeld, N., K.~Oshiba, Y.~Singer. 2020.
\newblock Predicting choice with set-dependent aggregation.
\newblock {\it International Conference on Machine Learning\/}. 2635--2644.

\bibitem[{Schwartz(2004)}]{ParadoxOfChoice}
Schwartz, Barry. 2004.
\newblock {\it The paradox of choice: {Why} more is less\/}.
\newblock HarperCollins, New York, NY, US.

\bibitem[{Seshadri et~al.(2019)Seshadri, Peysakhovich, and Ugander}]{CDM}
Seshadri, A., A.~Peysakhovich, J.~Ugander. 2019.
\newblock Discovering context effects from raw choice data.
\newblock {\it International Conference on Machine Learning\/}. 5660--5669.

\bibitem[{Simonson(1989)}]{attreffect1_simonson1989choice}
Simonson, I. 1989.
\newblock Choice based on reasons: The case of attraction and compromise
  effects.
\newblock {\it Journal of consumer research\/} {\bf 16}(2) 158--174.

\bibitem[{Simonson and Tversky(1992)}]{simonson1992choice_exp}
Simonson, Itamar, Amos Tversky. 1992.
\newblock Choice in context: Tradeoff contrast and extremeness aversion.
\newblock {\it Journal of marketing research\/} {\bf 29}(3) 281--295.

\bibitem[{Strauss et~al.(2018)Strauss, Klein, and
  Steinhardt}]{strauss2018review}
Strauss, A.~K, R.~Klein, C.~Steinhardt. 2018.
\newblock A review of choice-based revenue management: Theory and methods.
\newblock {\it European Journal of Operational Research\/} {\bf 271}(2)
  375--387.

\bibitem[{Talluri and Van~Ryzin(2004)}]{Talluri2004a}
Talluri, K.~T., G.~J. Van~Ryzin. 2004.
\newblock {\it The Theory and Practice of Revenue Management\/}.
\newblock Springer, Boston, MA, US.

\bibitem[{Thurstone(1927)}]{thurstone1927RUM}
Thurstone, L.~L. 1927.
\newblock A law of comparative judgment.
\newblock {\it Psychological review\/} {\bf 34}(4) 273.

\bibitem[{Tversky and Simonson(1993)}]{Tversky1993}
Tversky, A., I.~Simonson. 1993.
\newblock {Context-Dependent Preferences}.
\newblock {\it Management Science\/} {\bf 39}(10) 1179--1189.

\bibitem[{Usher and McClelland(2004)}]{Usher2004a}
Usher, M., J.~L. McClelland. 2004.
\newblock Loss aversion and inhibition in dynamical models of multialternative
  choice.
\newblock {\it Psychological review\/} {\bf 111}(3) 757.

\bibitem[{van Ryzin and Vulcano(2015)}]{Vulcano-marketDiscovery}
van Ryzin, G., G.~Vulcano. 2015.
\newblock {A Market Discovery Algorithm to Estimate a General Class of
  Nonparametric Choice Models}.
\newblock {\it Management Science\/} {\bf 61}(2) 281--300.

\bibitem[{Yellott(1977)}]{yellott1977uniformExpansion}
Yellott, J.~I. 1977.
\newblock The relationship between luce's choice axiom, thurstone's theory of
  comparative judgment, and the double exponential distribution.
\newblock {\it Journal of Mathematical Psychology\/} {\bf 15}(2) 109 -- 144.

\end{thebibliography}


\begin{APPENDIX}{}
\section{``The Economist'' Choice Experiment from  \citep{ariely2008predictably}}
\label{ap:econ}
\begin{exmp}
\label{exmp:econ}
\rev{
In this experiment, students where asked to choose among different subscription plans for the magazine ``The Economist''. In particular, the following three options were used : (1) Only version only, priced 50\$, (1) Printed version only, with a price of 125\$, and (3) Printed and Online subscription, at a price of 125\$. Table \ref{tab:marketShare_econ} reports the market share of the various options in choice scenarios $S_1$, where only options \{1,3\} were offered to students, and $S_2$, where students where able to choose among the three options \{1,2,3\}.  This experiment exhibits a violation of the regularity assumption, since the probability of choosing option (3) increases from 32\% to 84\% when option (2) is added to the offer set. Hence, no model belonging to the RUM class can perfectly fit this dataset.}

\begin{table}[!h]
    \centering
    \begin{tabular}{lr@{\extracolsep{1cm}}rr}
    \toprule
     &  & \multicolumn{2}{c}{Market Share} \\
     \cmidrule{3-4} 
     Versions & Price (\$)  & $S_1$ & $S_2$ \\
     \midrule
     (1) Online              &  $59$  & $.68$  & $ .16$\\
     (2) Printed             &  $125$ & $-$    & $ .0$\\
     (3) Printed \& Online   &  $125$ & $.32$  & $ .84$ \\
     \bottomrule
    \end{tabular}
    \caption{Predicted shares of three camera models in choice scenarios $S_1$, where respondents must choose between alternatives $\{1,2\}$, and $S_2$, where option $(3)$ is added to the offer set}
    \label{tab:marketShare_econ}
\end{table}

\end{exmp}

\rev{
The choice phenomena reported in Table \ref{tab:marketShare_econ} is an example of the so-called \textit{decoy effect}. In fact, option (2) is clearly ``dominated'' by option (3) in terms of attractiveness, which allows, for the same price, to obtain both the Printed \textit{and} Online versions of the magazine. Options perceived inferior in terms of quality and/or price, i.e., decoy options, are often used in Marketing to increase the perceived attractiveness of other products.  }

\rev{In Table \ref{tab:GSP_mod_econ} we report a GSP choice model perfectly fitting the choice outcome of the experiment in Example \ref{exmp:econ}. Specifically, we report the only three behaviors $C_k(\sigma_k, i_k)$, with $k=1,2,3$, with non-zero probability, and how choice are determined in the two choice scenarios. Finally, Table \ref{tab:PredShare_econ} reports the choice probabilities predicted by the model, which match the observed ones from Table \ref{tab:marketShare_econ}.}

\begin{table}[h!]
      \centering
        \begin{tabular}{ccc@{\extracolsep{0.7cm}}r@{\extracolsep{1cm}}c@{\extracolsep{0.5cm}}c}
        \toprule
            \multicolumn{3}{c}{Customer Type} &  Probability & $S_1=\{1,3\}$ & $S_2=\{1,2,3\}$ \\
        \cmidrule{1-3}
            $k$   & $\sigma_k$ & $i_k$ & $\lambda_k$ & $\sigma_{k,S_1}$  & $\sigma_{k,S_2}$\\
        \midrule
            $1$   &  $(3,1,2)$   &  1   & 0.16 & $(\pmb{3},1)$ & $(\pmb{3},1,2)$\\
            $2$   &  $(2,1,3)$   &  2   & 0.16 & $(1, \pmb{3})$ & $(2,\pmb{1},3)$\\
            $3$   &  $(2,3,1)$   &  2   & 0.68 & $(3, \pmb{1})$ & $(2,\pmb{3},1)$\\
        \bottomrule
        \end{tabular}
        \caption{GSP model from \cite{gsp-berbeglia2018} explaining the choice outcomes of Example \ref{exmp:econ}. For each $\sigma_{k,S}$, we highlight in bold the chosen item $j: \sigma_{k,S}(j) = i_k$.} 
        \label{tab:GSP_mod_econ}
\end{table}%

\begin{table}[!h]
    \centering
    \begin{tabular}{lr@{\extracolsep{1cm}}rr}
    \toprule
     &  & \multicolumn{2}{c}{Predicted Share} \\
     \cmidrule{3-4} 
     Versions & Price (\$)  & $S_1$ & $S_2$ \\
     \midrule
     (1) Online     &  $59$ & $\lambda_3 = .68$ & $\lambda_2 = .16$\\
     (2) Printed    &  $125$ & $-$ & $0$\\
     (3) Printed \& Online   &  $125$ & $\lambda_1 + \lambda_2 = .32$  & $\lambda_1 + \lambda_3=.84$ \\
     \bottomrule
    \end{tabular}
    \caption{Predicted shares of three camera models in choice scenarios $S_1$, where respondents must choose between alternatives $\{1,2\}$, and $S_2$, where option $(3)$ is added to the offer set}
    \label{tab:PredShare_econ}
\end{table}

\section{Regularity violation of the no-purchase option}
\label{ap:regularity0}
The Generalized Stochastic Preference choice model as defined by \cite{gsp-berbeglia2018} does not account for violations of the regularity assumption for the no-purchase option \citep[see][Lemma 1]{gsp-berbeglia2018}. Given two offer sets $S \subseteq S' \subseteq \mathcal{N}$, in particular, the authors show that every customer choosing the no-purchase option from $S'$, by definition, must choose the no-purchase option from $S$ as well. However, we can circumvent such limitation by allowing a customer type $C_k(\sigma_k, i_k)$ to rank the no-purchase option in $\sigma$. Consider, for example, two offer sets $S=\{0,1,2\}$ and $S'=\{0,1,2,3\}$, where $S \subset S'$, and a customer type $C_1\big((3~0~1~2), 1\big)$. Her choice behavior is reported in Table \ref{ap:eg:regularity0}. It is easy to see that, by introducing the option 3 in the offer set, we can increase the probability of option 0 being chosen and, thus, of the customer leaving without any purchase. 

\begin{table}[!htbp]
    \centering
    \begin{tabular}{lrr}
        \toprule
        \multicolumn{1}{c}{$S$} & \multicolumn{1}{c}{$\sigma_{1,S}$} & Choice \\
         \bottomrule
         $\{0,1,2\}$&   (0 1 2)   & 1\\
         $\{0,1,2,3\}$& (3 0 1 2) & 0\\
         \bottomrule
        \end{tabular}
    \caption{Choice behavior of customer $C\big((3~0~1~2), 1\big)$ faced with two different offer sets.}
    \label{ap:eg:regularity0}
\end{table}

\vspace{-0.5cm}


\section{Details on the implementation of PCMC}
\label{ap:pcmc_implem}
In this section, we elaborate on the implementation details of the PCMC choice model. The model is trained by Maximum Likelihood Estimation, and a Sequential Least SQuares Programming (SLSQP) solver \citep{nocedal2006numerical} is used to optimize the corresponding objective function, which is concave in general. The authors suggest to use additive-smoothing to avoid some numerical issues involved in the training of the model. In particular, given an offer set of size $|S|$ and an additive smoothing parameter $\alpha$, the probability of choosing alternative $j$ is computed at training time as

\begin{table}[]
  \centering
  \small
    \begin{tabular}{@{\extracolsep{4pt}}lrrrrr@{}}
    \toprule[1.2pt]
    \multicolumn{2}{c}{Irrational instances} & \multicolumn{2}{c}{PCMC -25} & \multicolumn{2}{c}{PCMC - $\infty$} \\
    \cmidrule{3-4} \cmidrule{5-6}
          & \multicolumn{1}{c}{\% irrat} & \multicolumn{1}{c}{Crossval} & \multicolumn{1}{c}{None} & \multicolumn{1}{c}{Crossval} & \multicolumn{1}{c}{None} \\
    \bottomrule[1.2pt]
   \T Halo-MNL(1) & 10   & \textbf{0.2595} & 0.2610 & 0.2897 & 0.2941  \\
          & 25  & \textbf{0.3756} & 0.3804 & 0.3933 & 0.3937 \\
          & \multicolumn{1}{l}{avg (all)} & \textbf{0.3175} & 0.3207 & 0.3415 & 0.3439 \\
    \midrule
    Halo-MNL(10) & 10   & \textbf{0.1997} & 0.2195 & 0.2348 & 0.2545 \\
          & 25  & \textbf{0.2430} & 0.2518 & 0.2873 & 0.3010 \\
          & \multicolumn{1}{l}{avg (all)} & \textbf{0.2214} & 0.2356 & 0.2610 & 0.2778 \\
    \midrule
    GSP(10) & 10   & 0.4293 & \textbf{0.4273} & 0.4615 & 0.4625 \\
          & 20   & \textbf{0.4622} & 0.4625 & 0.4927 & 0.4918 \\
          & 50   & \textbf{0.5283} & 0.5312 & 0.5668 & 0.5711 \\
          & \multicolumn{1}{l}{avg (all)} & \textbf{0.4733} & 0.4737 & 0.5070 & 0.5084 \\
    \midrule
    GSP(100) & 10   & \textbf{0.2790} & 0.2890 & 0.3223 & 0.3386 \\
          & 20   & \textbf{0.2880} & 0.3009 & 0.3354 & 0.3478 \\
          & 50   & \textbf{0.3283} & 0.3394 & 0.3868 & 0.3976 \\
          & \multicolumn{1}{l}{avg (all)} & \textbf{0.2984} & 0.3098 & 0.3481 & 0.3613 \\
    \midrule
    avg (all) &       & \textbf{0.3500} & 0.3568 & 0.3887 & 0.3967 \\
    \bottomrule[1.2pt]
    \multicolumn{2}{c}{Rational instances} &       &       &       & \T \\
    \bottomrule[1.2pt]
    \T MNL   & \multicolumn{1}{c}{-} & \textbf{0.1351} & 0.1404 & 0.1930 & 0.2012  \\
    MMNL  & \multicolumn{1}{c}{-} & \textbf{0.1381} & 0.1529 & 0.1970 & 0.2136 \\
    RB(10) & \multicolumn{1}{c}{-} & \textbf{0.3840} & 0.3841 & 0.4028 & 0.4065 \\
    RB(100) & \multicolumn{1}{c}{-} & \textbf{0.2741} & 0.2793 & 0.3167 & 0.3292 \\
    \midrule
    avg (all) &       & \textbf{0.2328} & 0.2392 & 0.2774 & 0.2876 \\
    \bottomrule[1.2pt]
    \end{tabular}%
  \caption{Average $L_1$ test errors for different PCMC implementations under various ground truth models. Each line averages over instances generated with different number of training offer sets (10,20 and 50) and transactions (3,000 and 50,000).}
  \label{ap:tab:pcmc:all versions:aggregated}%
\end{table}%

\begin{table}[]
  \centering
  \small
    \begin{tabular}{@{\extracolsep{4pt}}lrrrrr@{}}
    \toprule[1.2pt]
          \multicolumn{2}{c}{Irrational}        & \multicolumn{2}{c}{PCMC - 25} & \multicolumn{2}{c}{PCMC - $\infty$} \\
          \cmidrule{3-4} \cmidrule{5-6}
          &   $M$    & \multicolumn{1}{c}{Crossval} & \multicolumn{1}{c}{None} & \multicolumn{1}{c}{Crossval} & \multicolumn{1}{c}{None} \\
    \bottomrule[1.2pt]
    \T Halo-MNL & 10    & \textbf{0.3091} & 0.3097 & 0.3341 & 0.3332 \\
          & 20    & 0.2824 & \textbf{0.2801} & 0.2827 & 0.2826 \\
          & 50    & 0.2340 & 0.2329 & \textbf{0.1546} & 0.1570 \\
    \midrule
   GSP & 10    & \textbf{0.4859} & 0.4873 & 0.5347 & 0.5340 \\
          & 20    & \textbf{0.4052} & 0.4069 & 0.4300 & 0.4322 \\
          & 50    & 0.3355 & 0.3383 & \textbf{0.2817} & 0.2827 \\
    \midrule
    avg (all) &       & 0.3677 & 0.3688 & \textbf{0.3667} & 0.3675 \\
    \bottomrule[1.2pt]
    \multicolumn{2}{c}{Rational} &       &       &       & \T \\
    \bottomrule[1.2pt]
    \T (M)MNL & 10    & \textbf{0.1212} & 0.1216 & 0.1518 & 0.1497 \\
          & 20    & \textbf{0.0921} & 0.0924 & 0.1231 & 0.1294 \\
          & 50    & 0.0733 & 0.0704 & \textbf{0.0659} & 0.0662 \\
    \midrule
    RB & 10    & \textbf{0.4187} & 0.4291 & 0.4897 & 0.4997 \\
          & 20    & 0.3530 & \textbf{0.3429} & 0.3794 & 0.3893 \\
          & 50    & 0.3072 & 0.2961 & 0.2099 & \textbf{0.2086} \\
    \midrule
    all (avg) &       & \textbf{0.3490} & 0.3497 & 0.3494 & 0.3506 \\
    \bottomrule[1.2pt]
    \end{tabular}%
  \caption{Average $L_1$ test errors for different PCMC implementations under various ground truth models, on instances with 50,000 choice samples available for training. Instances are further divided based on the number of offer sets $M$ observed during training. }
  \label{ap:tab:pcmc50000trans}%
\end{table}%
\vspace{-0.3cm}

\begin{equation*}
    P(j|S) = \frac{T_{jS} + \alpha }{T_S + \alpha |S|},
\end{equation*}

\vspace{0.05cm}

\noindent
where $T_S$ is the number of training samples showing offer set $S$, and $T_{jS}$ the number of times alternative $j$ is chosen from the offer set $S$. In Table \ref{ap:tab:pcmc:all versions:aggregated}, we investigate the change in performance due to different stopping criteria and values of parameter $\alpha$. In particular, we implemented stopping criteria based on 
\begin{enumerate}
    \item The maximum number of iterations to be performed by the solver: this is set to $25$, which is the default value in the code provided by the authors. The corresponding results are reported in column ``PCMC-25''
    \item The absolute change in the objective function between two consecutive iterations: the algorithm is stopped when this change is smaller than $10^{-6}$, and the corresponding results are reported in column ``PCMC-$\infty$''.
\end{enumerate}
Moreover, for each stopping criterion, we compared the performance obtained by using different amounts of additive-smoothing in the training set. Specifically, column ``Crossval'' reports the average generalization error obtained using 5-folds crossvalidation to select the best $\alpha \in \{0, 0.01, 0.1, 1, 5, 10\}$. Column ``None'' corresponds to $\alpha=0$, for which no additive smoothing was used.
  
  The average $L_1$ test error has been reported over instances grouped by ground-truth models and number of customers types, indicated in parenthesis in the first column. The value of column ``\% irrat'' further divides each group based on the characteristics of the ground-truth model generating the corresponding set of instances. For Halo-MNL instances,  this column indicates the amount of pairwise interactions among products, while, for GSP instances, it indicates the percentage of irrational behaviors. We observe that limiting the number of iterations seems to be having a major impact on the predictive accuracy of the resulting choice model. Our intuition is that allowing the solver to proceed until convergence is reached may end up in overfitting the training set. We also notice that the gain in performance obtained by using additive-smoothing is not significant in general. To confirm whether the deterioration in performance of PCMC-$\infty$ is actually due to overfitting, in Table \ref{ap:tab:pcmc50000trans} we further compare the two variants on a set of instances where $50,000$ training samples have been generated, and we investigate the impact of the number of training offer sets on the resulting choice model. Confirming our previous hypothesis, we notice that when significant amount of training data is available, both in terms of number of training samples and number of training offer sets $M$, the risk of overfitting decreases and a better fit at training time translates in a significant improvement in generalization error. Nevertheless, also for this set of instances, i.e., with $M=50$ offer sets and $50,000$ samples are available at training time, the performance of the best PCMC variant is worse than the one of the GPT-based approaches. At this point, one may wonder whether more adaptive stopping criteria may be used instead of fixing ahead the maximum number of iterations. However, further experiments revealed that fixing the number of iterations to $25$ worked better on average than other stopping criteria based on
\begin{itemize}
 \item The relative change in the objective function between consecutive iterations ($< 1\%$),
 \item The maximum absolute change in the predicted probabilities over all training offer sets ($<0.001$),
 \item The maximum absolute change in the value of the parameters of the PCMC choice model,
 \item The maximum number of iterations set to $100$.
\end{itemize}

We thus avoid reporting the set of results corresponding to such stopping criteria, and use the PCMC-25 variant with no additive smoothing in the rest of our experiments. In particular, this is also the PCMC implementation used for the experiments reported in Section \ref{subsec:synthetic_exps} and Section \ref{subsec:real_exps}.  

\section{Additional Numerical Results}
\label{ap:add_res}
    \subsection{Learning statistics}
    \label{ap:learning_stats}
    
       \vspace{0.1cm}
     \begin{table}[!htpb]
         \centering
         \small
          
        \begin{tabular}{lrrrrrrrrrr}																																																																																
        \toprule																																																																																
        	Instances	&		& \multicolumn{3}{c}{Max ranked} &			\multicolumn{3}{c}{Max irrat level} &			\multicolumn{3}{c}{Prob. Irrat. Customer}												\\																																						\cmidrule(lr){3-5}	\cmidrule(lr){6-8}\cmidrule(lr){9-11}																			
        		&	\%irrat	&	GPT-R	&	GPT-I	&	GPT-IC	&	GPT-R	&	GPT-I	&	GPT-IC	&	GPT-R	&	GPT-I	&	GPT-IC	\\																																																	
        		\midrule
        	MNL	&	0	&	2.8	&	2.9	&	2.1	&	0	&	1.7	&	1.1	&	0	&	0.45	&	0.39	\\																																																										
        	Halo-MNL(1)	&	10	&	4.3	&	4.4	&	2.7	&	0	&	2.9	&	1.7	&	0	&	0.51	&	0.40	\\																																																										
        	Halo-MNL(1)	&	25	&	4.8	&	5.8	&	3.3	&	0	&	3.7	&	2.1	&	0	&	0.61	&	0.49	\\																																																										
        	\addlinespace																																																																															
        	MMNL	&	0	&	2.3	&	2.4	&	2.0	&	0	&	1.3	&	1.0	&	0	&	0.43	&	0.40	\\																																																										
        	Halo-MNL(10)	&	10	&	2.9	&	2.8	&	2.0	&	0	&	1.7	&	1.0	&	0	&	0.51	&	0.43	\\																																																										
        	Halo-MNL(10)	&	25	&	3.5	&	3.4	&	2.4	&	0	&	2.2	&	1.4	&	0	&	0.56	&	0.47	\\																																																										
        	\addlinespace																																																																														
        	RB(10)	&	0	&	4.8	&	5.5	&	3.7	&	0	&	3.7	&	2.2	&	0	&	0.34	&	0.30	\\																																																										
        	GSP(10)	&	10	&	4.9	&	5.6	&	3.7	&	0	&	3.8	&	2.3	&	0	&	0.40	&	0.35	\\																																																										
        	GSP(10)	&	20	&	4.9	&	6.1	&	3.9	&	0	&	4.2	&	2.5	&	0	&	0.45	&	0.38	\\																																																										
        	GSP(10)	&	50	&	5.0	&	6.9	&	4.1	&	0	&	4.9	&	2.8	&	0	&	0.56	&	0.45	\\																																																										
        	\addlinespace																																																																															
        	RB(100)	&	0	&	3.3	&	3.4	&	2.4	&	0	&	2.1	&	1.4	&	0	&	0.49	&	0.40	\\																																																										
        	GSP(100)	&	10	&	3.3	&	3.5	&	2.4	&	0	&	2.3	&	1.4	&	0	&	0.52	&	0.44	\\																																																										
        	GSP(100)	&	20	&	3.5	&	3.7	&	2.5	&	0	&	2.5	&	1.4	&	0	&	0.56	&	0.47	\\																																																										
        	GSP(100)	&	50	&	3.8	&	4.4	&	2.6	&	0	&	3.1	&	1.5	&	0	&	0.62	&	0.50	\\																																																										
        	\addlinespace																																																																															
        	(All) Mean	&	-	&		4.0	&	4.6	&	2.9	& 0	&	3.1	&	1.8	&	0	&	0.51	&	0.43	\\																																																										
        	(All) Median	&	-	&	4	&	4	&	3 	& 	0	&	3	&	2	&	0	&	0.51	&	0.43	\\																																																										
        	(All) Max	&	-	&		9	&	10	&	9	& 0	&	9	&	8	&	0	&	0.96	&	0.96	\\																																																										
        \bottomrule																																																																																
        \end{tabular}															
         \caption{Statistics describing choice models learned by GPT-based approaches.}
         \label{tab:Learning_stats}
     \end{table}
     \rev{
        We now elaborate on the impact of the irrationality level of an instance on the choice models learned by GPT-based approaches. In particular, Table \ref{tab:Learning_stats} groups instances based on ground-truth models and number of customer types used to generate the data. For each group, we sort Halo-MNL and GSP instances by increasing levels of irrational interactions and behaviors, respectively, in the ground-truth model. We then report, for each approach, its maximum number of strictly ranked products, the maximum irrationality level of its customer types, and the total probability of a customer being irrational. This, in particular, corresponds to the sum of the probabilities of irrational customer types in the learned choice model.}
        
        \rev{Confirming the results from \cite{jena2020partially}, all approaches strictly rank only a relatively small number of products on average, but are able to capture high order interactions when needed, by strictly ranking up to 10 products in specific cases. Also, for the irrational approaches, i.e., GPT-I and GPT-IC, both the maximum level of irrationality and the probability of a customer being irrational seem to generally increase with the level of irrationality of the ground-truth model. However, we note that on rational instances, GPT-I and GPT-IC still predict a customer being irrational with a relatively high probability. While, as observed in Section \ref{sec:lor}, high number of irrational customer classes may still result in close-to-rational behavior at the population level, this may justify the superior performance of GPT-R on such instances.
        }

      \begin{figure}[!htpb]
            \centering
            \includegraphics[scale=0.5]{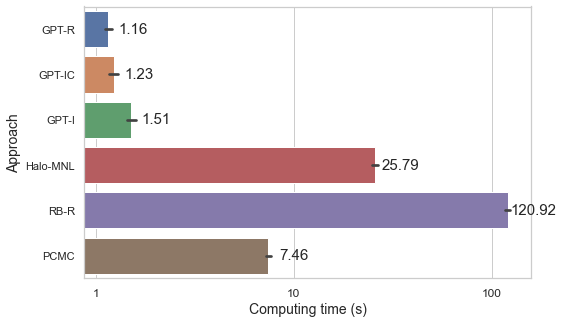}
            \caption{Average Computing times (seconds) for the various approaches over all synthetic instances.}
            \label{fig:comput_times}
        \end{figure}
        
        \rev{We conclude this set of analysis by diving into the computational aspect of the GPT-estimation procedure. In particular, as argued in Section \ref{sec:newrule}, the computational cost one has to pay for extending the algorithm from \cite{jena2020partially} to include irrational behaviors, stems from the fact that splitting each node (i.e., behavior) of the search-tree results in $|I(\sigma)| \cdot |P(\sigma)| \in O(N^2)$ new sub-behaviors, compared to the $|I(\sigma)| \in O(N)$ ones of the rational case. However, as observed from Table \ref{tab:Learning_stats}, the GPT procedure tends to discover customer types whose number of strictly ranked products $|P(\sigma)|$ is rather small. In practice, this makes the discovery of irrational behaviors computationally efficient. In this regard, we report in Figure \ref{fig:comput_times} the computing times of various approaches averaged over all synthetic instances, confirming the computational effectiveness of GPT-based approaches.}

    \vspace{0.1cm}
    \subsection{Impact of the amount of available data on the Loss of Rationality}
     \label{ap:dataImpact_lor}
     \rev{
     In this section we show that the Loss of rationality (LoR) can be influenced by factors other than the presence of irrational consumer behaviors. In particular, Figure \ref{fig:lor_data_av} shows that lower number of transactions available for training tend to lead to higher loss of rationality. Indeed, many interactions that may appear as irrational for small number of transactions, may be due to sampling noise and thus tend to disappear when more transactions become available. 
     
      \begin{figure}[!htpb]
     \begin{minipage}{0.5\textwidth}
         \centering
         \includegraphics[scale=0.55]{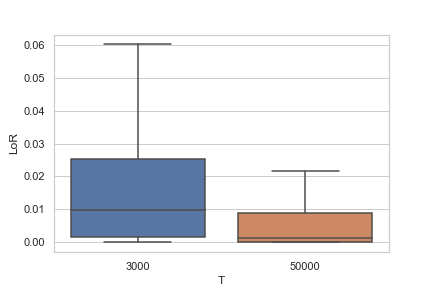}
     \end{minipage}
     \begin{minipage}{0.5\textwidth}
         \centering
         \includegraphics[scale=0.55]{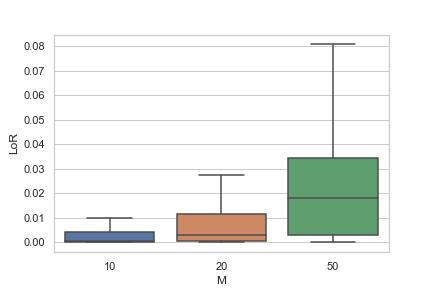}
     \end{minipage}
     \caption{Impact of the number of transactions $T$ (Left) and unique assortments $M$ (Right) available for training on the Loss of Rationality of instances. }
     \label{fig:lor_data_av}
     \end{figure}
     
     Another source of increase in the LoR of an instance stems from the number of unique assortments available for training. Indeed, consider the extreme case in which transactions are available for only one assortment. Such data, clearly, can be perfectly fit by any choice (or independent demand) model, since no substitution nor halo effects can be observed in this case. Hence, its LoR will be zero. As the number of assortments increases, more patterns of interactions among products become available, thus resulting in possibly complex choice probability distributions that may contain irrational interactions as well. Datasets with many training assortments are thus more likely to be associated with high LoR. 
     
    The two observations above shed light on the limitation of the LoR metric  as a tool for model selection, since no \textit{a priori} consideration can be made about an ``acceptable'' level of LoR beyond which one may consider to go beyond RUM. 
     
    }
    
\subsection{Impact of the irrationality level of GSP customer types.}
     \label{ap:gsp_inst}
     \rev{
     As observed in Section \ref{sec:newrule}, generalized stochastic preferences with different irrationality levels imply different types of interaction among products. We are thus interested in understanding how well each of the approach we implemented generalizes under different levels of customer irrationality $i_{max}$. To this end, in Table \ref{tab:RB-10_breakdown} we group GSP instances by the number of customer types (in parenthesis) and maximum irrationality level $i_{max}$ in the ground-truth model. While particularly high levels of irrationality may not be very common in practice, they allow us to analyze possible limitations of approaches, given the limited irrationality assumption intrinsic to the GPT-procedure. }
    \begin{table}[!htpb]
        \centering
        \small
        \begin{tabular}{lrrrrrrrr}																	
        \toprule																	
        Instances	&	$i_{max}$	&	LoR	&	RB-R	&	GPT-R	&	GPT-I	&	GPT-IC	&	PCMC	&	Halo-MNL	\\
        \midrule																	
        GSP(10)	&	1	&	0.0332	&	0.3131	&	0.2963	&	\textbf{0.2601}	&	0.2819	&	0.4824	&	0.6923	\\
        GSP(10)	&	5	&	0.0438	&	0.3414	&	0.3261	&	\textbf{0.3145}	&	0.3297	&	0.5208	&	0.7659	\\
        GSP(10)	&	9	&	0.0260	&	0.2436	&	\textbf{0.2261}	&	0.2351	&	0.2458	&	0.4177	&	0.6645	\\
        	&	All (Mean)	&	0.0343	&	0.2994	&	0.2828	&	\textbf{0.2699}	&	0.2858	&	0.4736	&	0.7075	\\
        	\addlinespace
        GSP(100)	&	1	&	0.0043	&	0.1903	&	0.1654	&	0.1630	&	\textbf{0.1606}	&	0.3052	&	0.2748	\\
        GSP(100)	&	5	&	0.0057	&	0.1939	&	\textbf{0.1773}	&	0.1808	&	0.1806	&	0.3307	&	0.3006	\\
        GSP(100)	&	9	&	0.0060	&	0.1798	&	\textbf{0.1598}	&	0.1665	&	0.1642	&	0.2933	&	0.2840	\\
        	&	All (Mean)	&	0.0053	&	0.1880	&	\textbf{0.1675}	&	0.1701	&	0.1685	&	0.3098	&	0.2865	\\
        	\addlinespace
        All (Mean)	&		&	0.0198	&	0.2437	&	0.2252	&	\textbf{0.2200}	&	0.2272	&	0.3917	&	0.4970	\\
        All (Median)	&		&	0.0057	&	0.2142	&	0.1898	&	\textbf{0.1862}	&	0.1918	&	0.3634	&	0.3796	\\
        All (Max)	&		&	0.2343	&	0.9454	&	0.7877	&	0.8087	&	\textbf{0.7835}	&	1.0035	&	1.5872	\\
        \bottomrule																	
        \end{tabular}																	
        \caption{Test errors comparison on GSP instances grouped based on the irrationality level of customer types in the ground-truth model. The metric reported is the average L1 error per offer set}
  \label{tab:RB-10_breakdown}%
    \end{table}
    \rev{
    We start by noticing that, on average, both Halo-MNL and PCMC are outperformed by rank-based approaches on this set of experiments. Also, as observed in Section \ref{subsubsec:genperf_synth}, GPT-IC tends to dominate GPT-I on instances with several customer behaviors, while the opposite is true for instances with less number of customer behaviors. On such instances, GPT-IC still outperforms GPT-R for $i_{max}$ of 1, confirming the fact that GPT-IC is more suited to capture low-order interactions. }
    
    
    It is also interesting to note a decrease in the Loss of Rationality of instances generated for $i_{max}=9$ and, coherently, an improvement in the predictive accuracy of rational rank-based methods on the same set of instances. Indeed, given an offer set $S$ and a generalized stochastic preference $C_k(\sigma_k, 9)$, it is often the case that $|\sigma_{k,S}|<i$.
    We recall from Section \ref{sec:model} that, in such cases, the considered customer type leaves with no purchase. Intuitively, such a behavior can be more easily approximated by a rational choice model imposing a high probability mass on the no-purchase option. This also translates into predictions that are more accurate on average than those obtained for smaller levels of irrationality $i_{max}$.
    

    \vspace{0.2cm}
    


\subsection{Impact of the type of positive interaction}
\label{ap:halo_SymmVsAsymm}

\rev{We now investigate the difference in performance of the various approaches on Halo-MNL instances with symmetric versus asymmetric product interactions. The former scenario aims to represent the case of complementarity effects, where two products increase each other attractiveness when present in the offer set. This is the case, for example, of pasta and tomato sauce, or pancake mix and maple syrup. The latter aims to represent the decoy effect (see Appendix \ref{ap:econ}), where an item is included in the offer set for the purpose of making another item, perceived as much better, more attractive to the customer.}

\rev{
In Table \ref{tab:symmVSasymm} we reports the test errors of the each approach on classes of instances grouped by ground-truth model, number of customer types (in parenthesis), and the ``Type'' of interaction between items, i.e., Symmetric or Asymmetric ones. On average, GPT-IC is the best performing approach among the implemented ones. In fact, as observed in Section \ref{subsubsec:genperf_synth}, the Halo-MNL choice model tends to need significant amount of data to well approximate the data-generating model. We note that, even in the case of Halo-MNL instances with one customer type, Halo-MNL seems to struggle for Asymmetric interactions in particular. This is probably due to the fact that, given its $N^2$ parameters, Halo-MNL tends to recover less sparse matrices of pairwise interactions, especially when only limited amount of data is available. 
}

\begin{table}[!htpb]
    \centering
    \small
    \begin{tabular}{llrrrrrrr}																			
    \toprule																			
    		Instances	&	Type	&	LoR	&	RB-R	&	GPT-R	&	GPT-I	&	GPT-IC	&	PCMC	&	Halo-MNL	\\
    \midrule																			
    		Halo-MNL(1)	&	Asymm.	&	0.0117	&	0.2293	&	0.2378	&	0.1851	&	\textbf{0.1806}	&	0.2533	&	0.2060	\\
    		Halo-MNL(1)	&	Symm.	&	0.0229	&	0.3112	&	0.3207	&	0.2685	&	0.2643	&	0.3878	&	\textbf{0.2337}	\\
    			&	All (Mean)	&	0.0173	&	0.2703	&	0.2792	&	0.2268	&	0.2224	&	0.3205	&	\textbf{0.2199}	\\
    			\addlinespace
    		Halo-MNL(10)	&	Asymm.	&	0.0037	&	0.1508	&	0.1255	&	0.1125	&	\textbf{0.1095}	&	0.2145	&	0.1927	\\
    		Halo-MNL(10)	&	Symm.	&	0.0042	&	0.1643	&	0.1447	&	0.1291	&	\textbf{0.1263}	&	0.2568	&	0.2028	\\
    			&	All (Mean)	&	0.0040	&	0.1575	&	0.1351	&	0.1208	&	\textbf{0.1179}	&	0.2357	&	0.1977	\\
    			\addlinespace
    		(All) Mean	&		&	0.0106	&	0.2139	&	0.2072	&	0.1738	&	\textbf{0.1702}	&	0.2781	&	0.2088	\\
    		(All) Median	&		&	0.0031	&	0.1954	&	0.1839	&	0.1495	&	\textbf{0.1432}	&	0.2576	&	0.1943	\\
    		(All) Max	&		&	0.1030	&	0.4963	&	0.5055	&	0.4790	&	\textbf{0.4436}	&	0.7093	&	0.9779	\\
    \bottomrule																			
    \end{tabular}																			
    \caption{Test errors comparison on GSP instances grouped based on the irrationality level of customer types in the ground-truth model. The metric reported is the average L1 error per offer set}
    \label{tab:symmVSasymm}
\end{table}

\section{Proof of Observation 1}
\label{ap:proofObs1}

\rev{
Observation 1 states the number of spurious positive interactions of an irrational partially-ranked preference sequence $C(P(\sigma),I(\sigma),i)$. can be as high as $\sum\limits^{|P(\sigma)|-1}_{j=i-1} {j \choose i-1}$.


For illustrative purpose, note here that the strictly ranked preference list of $C(P(\sigma),I(\sigma),i)$ is a sequence $(\sigma^{-1}(1), \sigma^{-1}(2), \ldots,\sigma^{-1}(|P(\sigma)|))$, where $\sigma^{-1}(r)$ denotes the item ranked at position $r$ in $P(\sigma)$.

Depending on the offer set, any item in $P(\sigma)$ (except for the $i-1$ first items, which will never be chosen due to irrationality level $i$) may be positively influenced by items ranked higher in $P(\sigma)$. Summing over all items in $P(\sigma)$ that may be subject to spurious positive interactions, we have a total of $\sum\limits^{|P(\sigma)|}_{r=i} {SI_r}$ possible spurious positive interactions in $C(P(\sigma),I(\sigma),i)$, where $SI_r$ is the number of possible spurious interactions for item $\sigma^{-1}(r)$, ranked at position $r$.

Further, an item $\sigma^{-1}(r)$, ranked at position $r$ can be positively influenced by degree $i$ in a given offer set $S$ only when exactly $i-1$ of all items that have ranks smaller than $r$ (i.e., $(\sigma^{-1}(1), \sigma^{-1}(2),  \ldots,\sigma^{-1}(r-1))$) are present in offer set $S$. Among those $r-1$ items, there are exactly $r-1 \choose i-1$ combinations how to select $i-1$ items. We therefore can compute $SI_r = {r-1 \choose i-1}$.
The result follows as the total number of possible spurious positive interactions is $\sum\limits^{|P(\sigma)|}_{r=i} {{r-1 \choose i-1}}$.

\qed
}

\end{APPENDIX}{}

\end{document}